\documentclass[twocolumn,tighten]{aastex631}

\shorttitle{Tossing BH spin axes}
\shortauthors{Tauris}

\graphicspath{{./}{figures/}}

\begin{document}

\title{Tossing Black Hole Spin Axes}

\correspondingauthor{Thomas M. Tauris}
\email{tauris@mp.aau.dk}

\author[0000-0002-3865-7265]{Thomas M. Tauris}
\affiliation{Department of Materials and Production, Aalborg University, Skjernvej 4A, DK-9220~Aalborg {\O}st, Denmark}

\begin{abstract}
The detection of double black hole (BH+BH) mergers provides a unique possibility to understand their physical properties and origin. 
To date, the LIGO-Virgo-KAGRA network of high-frequency gravitational wave observatories have announced the detection of more than 85~BH+BH merger events \citep{aaa+22a}.
An important diagnostic feature that can be extracted from the data is the distribution of effective inspiral spins of the BHs.
This distribution is in clear tension with theoretical expectations from both 
an isolated binary star origin, which traditionally predicts close-to aligned BH component spins \citep{kal00,fsm+17}, and formation via dynamical interactions in dense stellar environments that predicts a symmetric distribution of effective inspiral spins \citep{mo10,rzp+16}.
Here it is demonstrated that isolated binary evolution can convincingly explain the observed data if BHs have their spin axis tossed during their formation process in the core collapse of a massive star, similarly to the process evidently acting in newborn neutron stars. BH formation without spin-axis tossing, however, has difficulties reproducing the observed data --- even if alignment of spins prior to the second core collapse is disregarded. 
Based on simulations with only a minimum of assumptions, constrains from empirical data can be made on the spin magnitudes of the first- and second-born BHs, thereby serving to better understand massive binary star evolution prior to the formation of BHs.
\end{abstract}


\section{Introduction} \label{sec:intro}
Gravitational wave (GW) astronomy began seven years ago with the discovery of the first binary black hole (BH+BH) merger \citep[GW150914;][]{aaa+16a}. 
Since then, more than 85 such BH+BH merger events have been reported \citep{aaa+22a}, besides mergers of double neutron stars (NS+NS) and mixed BH+NS events \citep{aaa+17c,aaa+21b}. 
The merger events of BH/NS binaries are the most powerful energy sources known in the Universe and bring new opportunities of insight to fundamental aspects of physics, including: new tests of gravity in a highly relativistic regime \citep{aaa+20f}, creation of energetic bursts of electromagnetic radiation \citep{aaa+17f,aaa+18c} and production of heavy chemical elements which decay and power an optical transient \citep{scj+17,cfk+17,dps+17,whs+19,met19}.
Their origin remains controversial and has sparked a plethora of new research on massive binary star evolution in order to understand the formation process of such compact object mergers \citep[e.g.][]{mlp+16,rcr16,md16b,ktl+16,zsr+19,mrf+19,bfq+20,bkf+20,knic20}.

\begin{figure*}
\centering
\vspace*{-0.5cm}\hspace*{+0.8cm}
\includegraphics[width=1.0\textwidth]{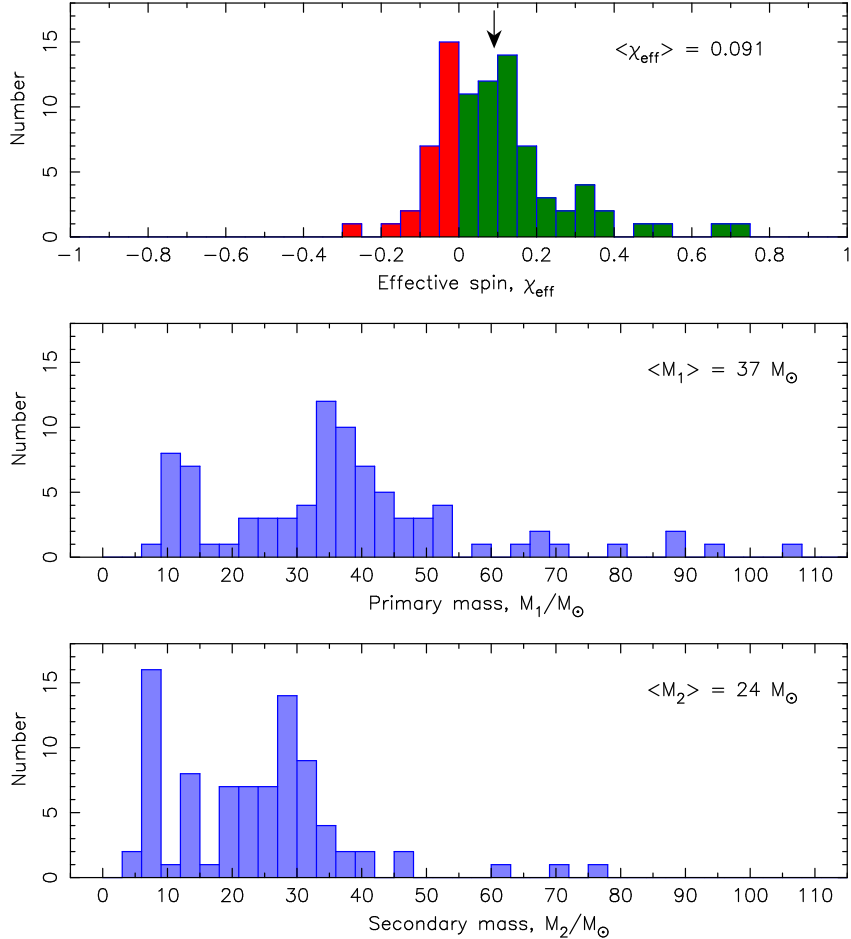} 
\caption{Top: Distribution of measured individual mean values of effective inspiral spins of 85 observed BH+BH merger events. Green and red colors indicate $\chi_{\rm eff}>0$ and $\chi_{\rm eff}<0$, respectively, and the arrow marks the average value, $\langle \chi_{\rm eff} \rangle = 0.091$. The distribution is clearly asymmetric and has a number of negative values. Central and bottom: Inferred component masses of primary and secondary BHs ($M_2>3.0\;M_\odot$) with average values stated.
Data is taken from GWTC-3 \citep{aaa+22a}. \label{fig:GWTC3_BHBH}}
\end{figure*}

The detected BH+BH mergers reveal individual BH component masses in the range from a few solar masses ($M_\odot$) to more than $100\;M_\odot$, thereby challenging both of the hypothesized mass gaps: at the lower and upper end of the BH mass spectrum, respectively \citep{aaa+21c,bjco98,rfj+20}. 
The other fundamental parameter describing an astrophysical BH is spin. The dimensionless spin magnitude is given by $\chi=cJ/GM^2$, 
where $J$ is the BH spin angular momentum, $M$ is its mass, and $c$ and $G$ are the speed of light and the gravitational constant, respectively. 

Unfortunately, it is difficult to directly measure the magnitudes of the individual BH spins ($\chi_1$ and $\chi_2$), or their tilt (misalignment) angles ($\Theta_{1,2}$) with respect to the orbital angular momentum vector of the binary ($\vec{L}$) during the inspiral of the merger. Only the sum of their projected spins along the orbital angular momentum is well measurable, i.e. the effective inspiral spin:\\
\\
\\
\begin{eqnarray}\label{eq:chi_eff2}
                 \chi _{\rm eff} & \equiv & \frac{\left(M_1\vec{\chi}_1+M_2\vec{\chi}_2\right)}{M_T} \cdot \frac{\vec{L}}{|\vec{L}|} \nonumber \\
                 & = & \frac{\chi_1 \cos \Theta_1 + q\,\chi_2\cos\Theta_2}{1+q}\,,
\end{eqnarray}
where $M_T = M_1 + M_2$ is the total mass of the two BHs and $q\equiv M_2/M_1\le 1$ is their mass ratio.
Whereas misaligned BH spins are subject to precession, $\chi_{\rm eff}$ is constant during the long inspiral prior to detection \citep{gks+15} and therefore an important diagnostic tool for the origin of BH+BH systems. 

Figure~\ref{fig:GWTC3_BHBH} (top panel) shows the distribution of measured individual mean values of $\chi_{\rm eff}$ of all BH+BH mergers and represents a conundrum due to a combination of two characteristics: i) the distribution is asymmetric \citep{aaa+20e,zbb+21,rco+21,aaa+22b}; and ii) there is apparently a significant number of systems having $\chi _{\rm eff}<0$ (although caution should be taken for the accuracy of individual values of $\chi_{\rm eff}$, see Appendix~\ref{appendix:A}). One the one hand, a perfect symmetric distribution of $\chi_{\rm eff}$ centered on $\chi_{\rm eff}=0$ would be expected if all events originated from binaries that assemble pairs of BHs with random individual spin directions \citep[e.g. via dynamical interactions in globular clusters,][]{mo10,rzp+16}. An asymmetric $\chi_ {\rm eff}$ distribution therefore points to, at least, a substantial contribution from BH+BH systems produced in isolated binaries. On the other hand, according to the current school of thought, negative effective spins (requiring at least one misalignment angle component $\Theta_i>90^{\circ}$) can only result if very large asymmetric momentum kicks are imparted onto BHs in their formation process \citep{wgo+18,cfr21,flr21}, following the collapse of a massive stellar core --- see Section~\ref{subsubsec:negative_spins}. However, large BH kicks are incompatible with observations of Galactic BH binary systems \citep{man16,mir17}. 
The only other way to produce $\chi_{\rm eff}<0$ is to prevent alignment of stellar spins prior to the formation of the second-born BH (see, however, Section~\ref{subsec:BH_component_spins}). 

Finally, it should be mentioned that BH+BH mergers could also have an origin from primordial BHs, population~III BH binaries, or dynamical assembly in AGN disks \citep[for a review, see e.g.][and references therein]{mf22}. However, in those scenarios the expected $\chi_{\rm eff}$ distribution is less clear.
For further discussions on interpretation of the empirical $\chi_{\rm eff}$ data, see e.g. \citet{sbm17,qfm+18,bfq+20,ob21,cmcf22}.

Here a different mechanism is investigated in which the formation of a BH in a supernova (SN) is accompanied by the tossing of its spin axis in a random (isotropic) direction. 
There is unambiguous observational evidence from high-precision timing of binary radio pulsars \citep{bkk+08,dkl+19} clearly demonstrating that such a process of spin-axis tossing is at work during the formation of NSs \citep{tkf+17}. In fact, in two out of two NS+NS systems in which the young NS is observable as a radio pulsar, its misalignment angle has been measured to be very significant: $50\pm0.4^\circ$ and $104\pm9^\circ$, respectively\footnote{For PSR~J0737$-$3039B there is an ambiguity of $180^\circ$, i.e. the misalignment angle could also be $\delta=130\pm0.4^\circ$ \citep{bre09}. Important here is that in either case $\delta \gg 0^\circ$.} \citep{bkk+08,dkl+19}.
Further evidence for tossing of the BH spin axis during the SN comes from recent measurements of a large spin-orbit misalignment ($>40^\circ$) in the X-ray binary MAXI~J1820+070 \citep{pvb+22}.
Hence, although the process responsible for the tossing of the BH spin axis following stellar core collapse remains unsettled (Section~\ref{subsubsec:mechanism}), there are accumulating pieces of evidence that the process of spin-axis tossing is a generic process in all core collapse SNe (although, as we shall discuss, the mechanism may be different for NSs and low-mass BHs, formed via fallback of SN ejecta, compared to massive BHs formed out of direct core collapse).

In Section~\ref{sec:SN_kinematics}, analytical SN kinematics is introduced for post-SN orbits without BH tossing, and sample results of BH+BH binaries are given. In Section~\ref{sec:simulations}, the applied method of Monte Carlo simulations with and without BH tossing is presented, along with a description of the chosen range of input parameter values and their probability distribution functions (PDFs). The results of these simulations, and comparison to BH+BH observational data, are presented and discussed in Section~\ref{sec:results}. Additional discussions are given in Section~\ref{sec:discussions} and the conclusions are summarized in Section~\ref{sec:conclusions}.

\section{Supernova kinematics}\label{sec:SN_kinematics}
\subsection{Analytical description}
The dynamical effects of SNe in close binaries have been studied in detail both analytically and numerically in a number of papers \citep[e.g.][]{fv75,hil83,bp95,kal96,tt98,tv23}. Here, we follow the prescriptions of \citet{hil83,tv23} and consider the explosion of a star (a stripped helium star of mass, $M_{\rm He}$) in a circular binary, producing the second-born BH with a mass, $M_{\rm BH,2}$. The companion star is in all cases assumed to be the first-born BH with a mass, $M_{\rm BH,1}$. 
The change in the orbital semi-major axis (i.e. the ratio of final post-SN to initial pre-SN value) can be expressed by:
\begin{equation}\label{eq:a_ratio_SN}
  \frac{a_{\rm f}}{a_{\rm i}}= \left[\frac{1-\Delta M/M}{1-2\Delta M/M -(w/v_{\rm rel})^2 -2\cos\theta \,(w/v_{\rm rel})} \right] \;,
\end{equation}
where $M=M_{\rm He}+M_{\rm BH,1}$ is the total mass of the pre-SN system and $v_{\rm rel}=\sqrt{G(M_{\rm He}+M_{\rm BH,1})/a_{\rm i}}$ is the relative velocity between the two stars. 
The kick angle, $\theta$ is defined as the angle between the kick velocity vector, $\vec{w}$ and the
pre-SN orbital velocity vector of the exploding star, $\vec{v}_{\rm He}$ in the centre-of-mass rest frame. 
Assuming a circular pre-SN binary is a good approximation for close binaries given the tidal interactions \citep{zah77,hut81} during Roche-lobe overflow (RLO) prior to the SN. 
Solving for the denominator being equal to zero in Eq.~(\ref{eq:a_ratio_SN}) yields the critical angle, $\theta _{\rm crit}$,
so that $\theta < \theta _{\rm crit}$ will result in a disruption of the orbit.

The eccentricity of the post-SN system can be evaluated directly from the post-SN orbital angular momentum, $L_{\rm orb,f}$ and  is given by:
\begin{equation}\label{eq:ecc} 
  e=\sqrt{1+\frac{2\,E_{\rm orb,f}\,L_{\rm orb,f}^2}{\mu_{\rm f}\,G^2 M_{\rm BH,1}^2 M_{\rm BH,2}^2}} \;,
\end{equation} 
where
\begin{equation} 
  L_{\rm orb,f}=a_{\rm i}\,\mu_{\rm f}\,\sqrt{\left(v_{\rm rel}+w\cos\theta\right)^2+\left(w\sin\theta\sin\phi\right)^2} \;.
\end{equation} 
Here $\mu_{\rm f}$ and $E_{\rm orb,f}=-GM_{\rm BH,1}M_{\rm BH,2}/2a_{\rm f}$ are the post-SN reduced mass and orbital energy, respectively. 
For a definition of the kick angles, $\theta$ and $\phi$, see fig.~13 in \citet{tkf+17}.

If the kick applied to the second-born BH is directed out of the orbital plane of the pre-SN system (i.e. if $\phi \ne \{0^\circ, \pm 180^\circ \}$), then the spin axis of the companion star (first-born BH) will be tilted with respect to 
post-SN orbital angular momentum vector. 
This effect gives rise to the geodetic precession seen in several radio pulsar binaries (in which there has been no mass transfer since the SN explosion) where the spin axis of the main-sequence star, white dwarf (WD) or NS companion is measured to be tilted \citep{kbm+96,kbv+20,ksm+06,fsk+13}. This misalignment angle can be calculated as: 
\begin{equation}\label{eq:misalign} 
   \delta = \cos ^{-1} \left( \frac{v_{\rm rel}+w\cos\theta}{\sqrt{(v_{\rm rel}+w\cos\theta)^2+(w\sin\theta\sin\phi)^2}} \right)\;.
\end{equation}
If the misalignment angle is large ($\delta > 90^\circ$), then the new orientation of the orbit will cause retrograde spin of one or both binary components to the sense of orbital revolution. It is trivial to see that a second critical kick angle, $\theta_{\rm retro}=\cos ^{-1}(-v_{\rm rel}/w)$ exists such that post-SN orbits will be retrograde if $\theta > \theta_{\rm retro} > 90^\circ$ (independent of $\phi$, see below).

\subsection{Analytical results}\label{subsec:analytical_results}
\begin{figure}
\vspace{-0.7cm}\hspace{-0.6cm}
\includegraphics[width=0.55\textwidth]{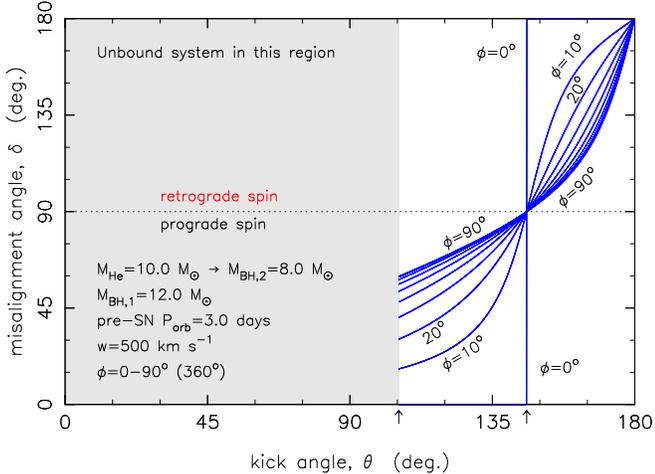} 
\caption{Misalignment angle, $\delta$ resulting from the formation of the second BH in a BH+BH system, for a fixed pre-SN orbital period of $P_{\rm orb}=3.0\;{\rm days}$ and a kick velocity of $w=500\;{\rm km\,s}^{-1}$, as a function of the two kick angles: $\theta$ and $\phi$. The mass of the collapsing star is assumed to be $M_{\rm He}=10.0\;M_\odot$ and it produces a BH of mass, $M_{\rm BH,2}=8.0\;M_\odot$. The mass of the companion star, the first-born BH, is assumed to be $M_{\rm BH,1}=12.0\;M_\odot$. The grey region to the left represents post-SN systems that are disrupted due to the SN, i.e. $\theta < \theta_{\rm crit}=105.4^{\circ}$. Above another critical angle, $\theta > \theta_{\rm retro}=145.8^{\circ}$, the post-SN orbits are always retrograde ($\delta > 90^{\circ}$, i.e. the new-formed BH is shot into an orbit such that the spin vector of the (old) first-born BH is pointing in the hemisphere opposite to that of the new orbital angular momentum vector). The blue lines represent values of $\phi$ between $0-90^\circ$, in steps of $10^\circ$. 
\label{fig:misalignment_kickangles}}
\end{figure}

\begin{figure}
\vspace{0.0cm}\hspace{-0.4cm}
\includegraphics[width=0.49\textwidth]{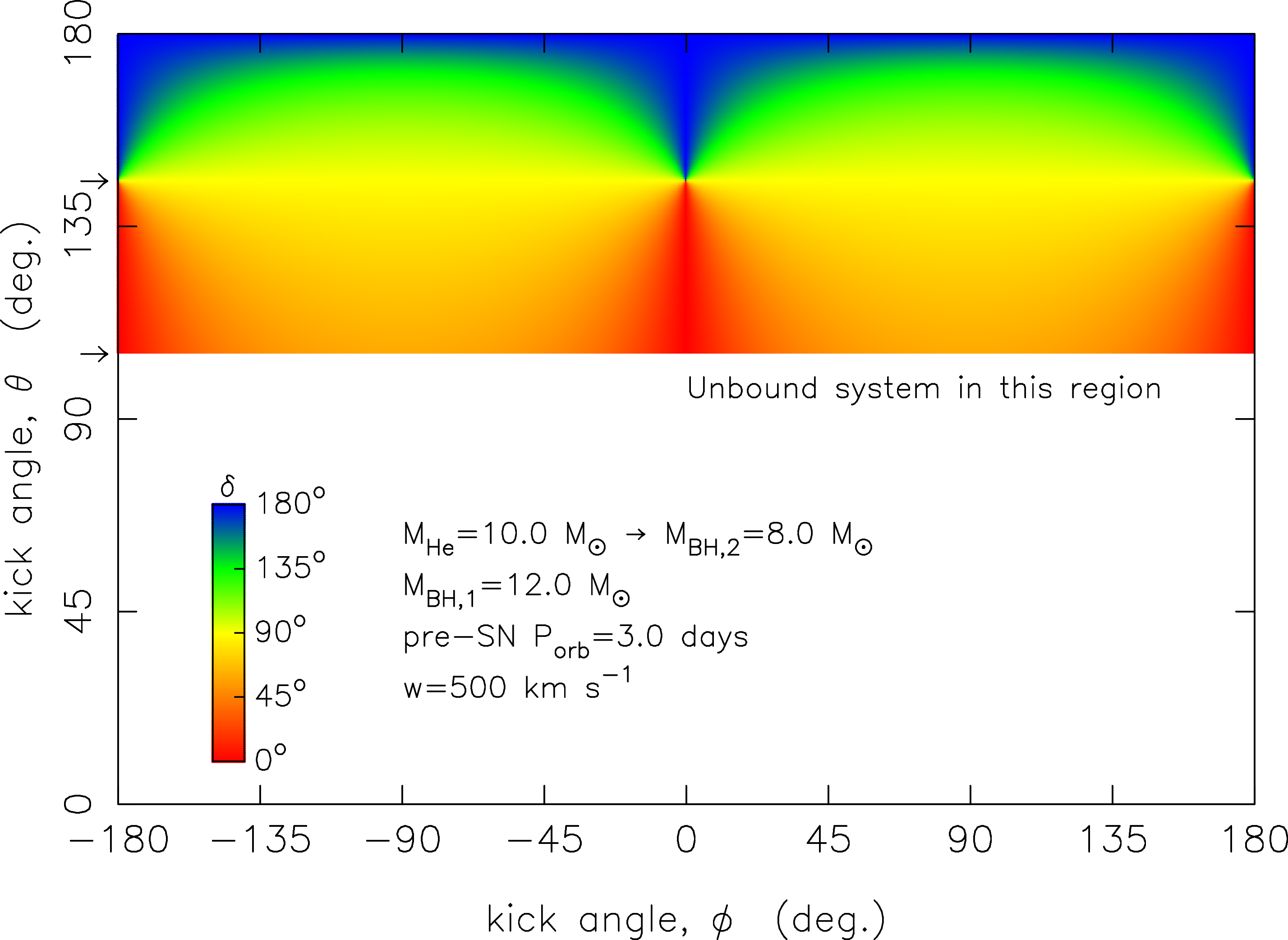}
\caption{Misalignment angle, $\delta$ (color coded; see vertical bar) as a function of the two kick angles, $\phi$ and $\theta$. The pre-SN parameters are identical to those of Fig.~\ref{fig:misalignment_kickangles}. The two angles $\theta_{\rm crit}=105.4^{\circ}$ and $\theta_{\rm retro}=145.8^{\circ}$, marked with arrows, are clearly noticeable (see also Fig.~\ref{fig:misalignment_kickangles}).
\label{fig:misalignment_theta-phi}}
\end{figure}

\begin{figure*}
\includegraphics[width=0.50\textwidth]{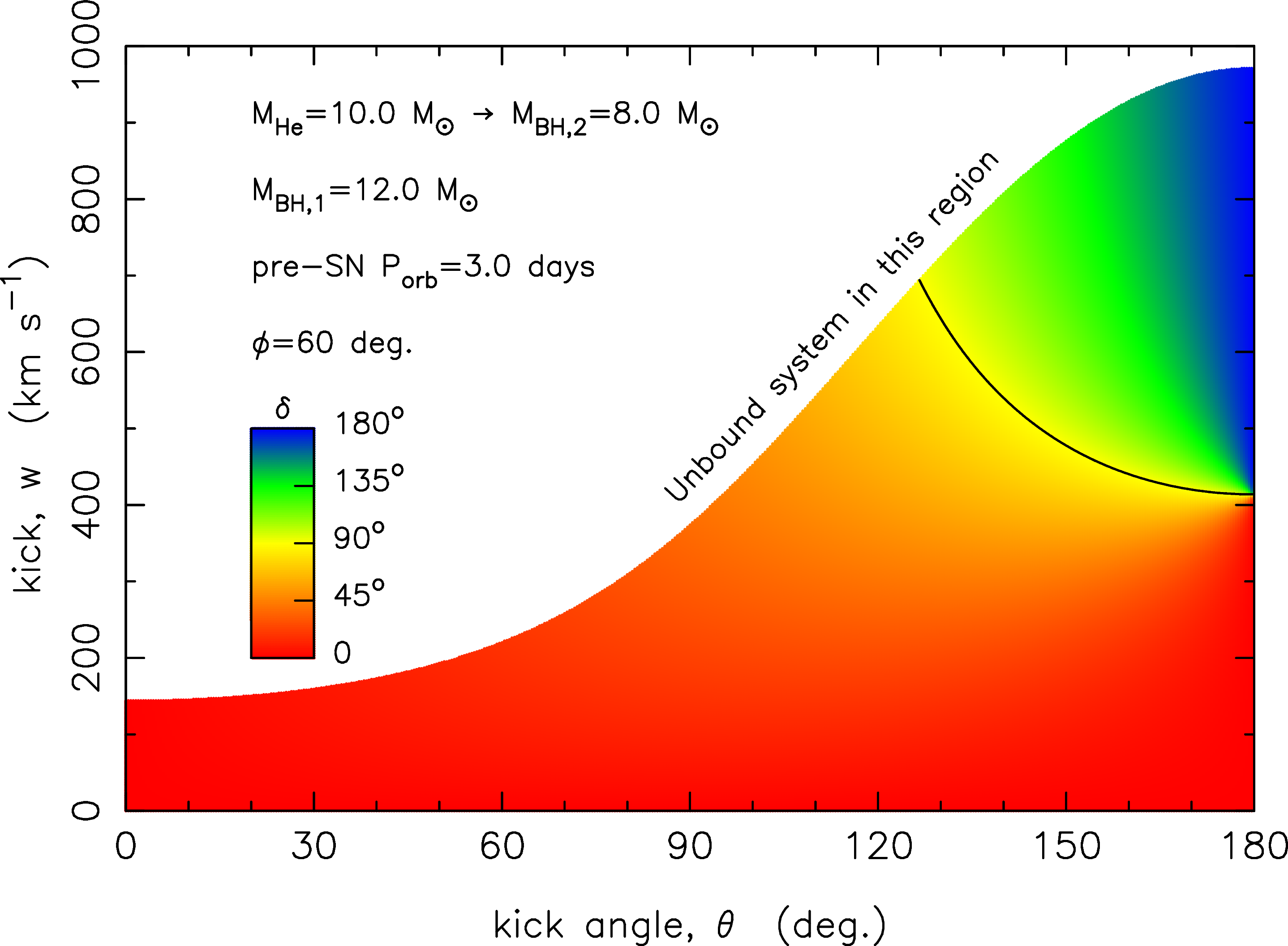}
\includegraphics[width=0.50\textwidth]{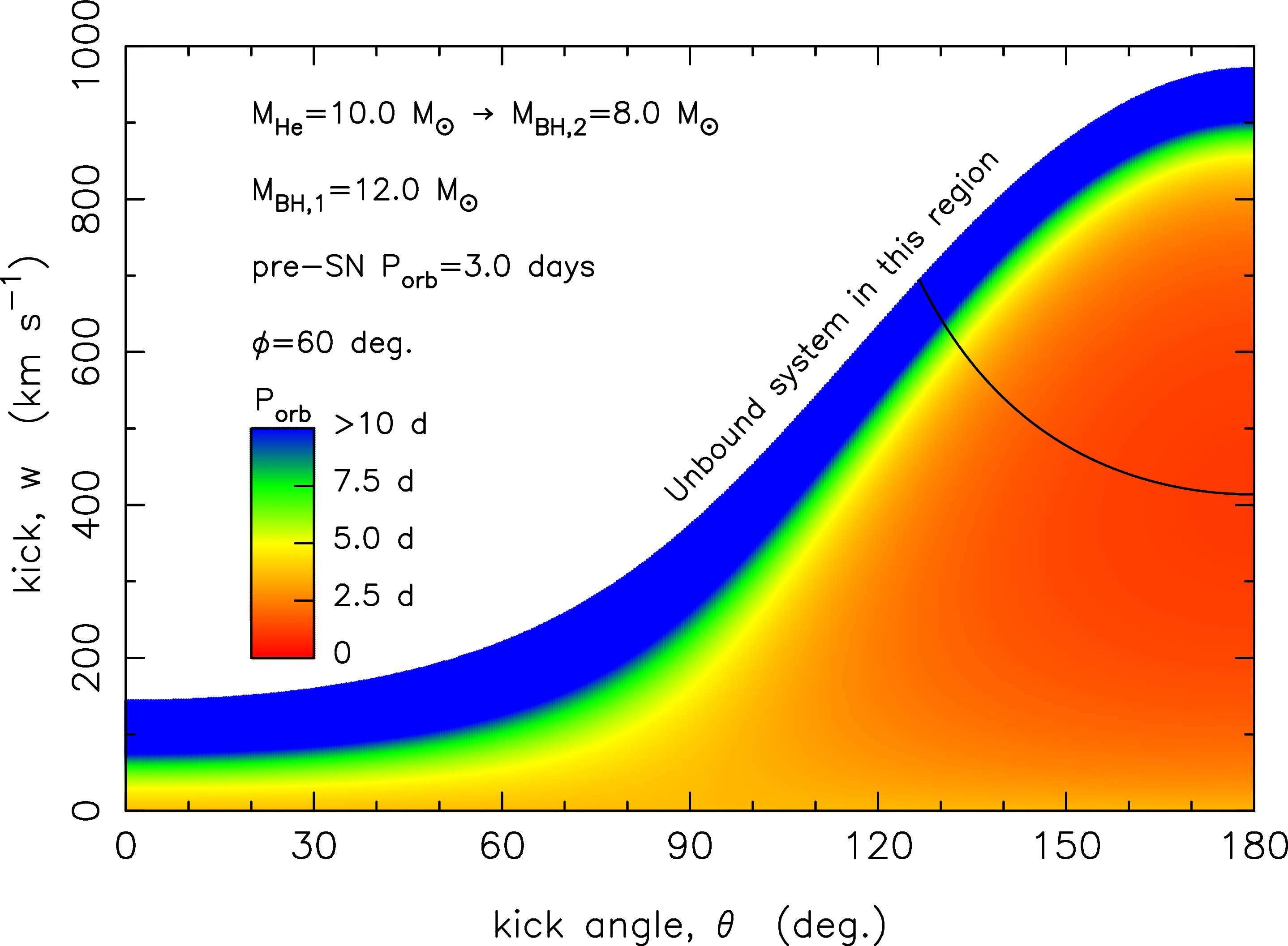}
\includegraphics[width=0.50\textwidth]{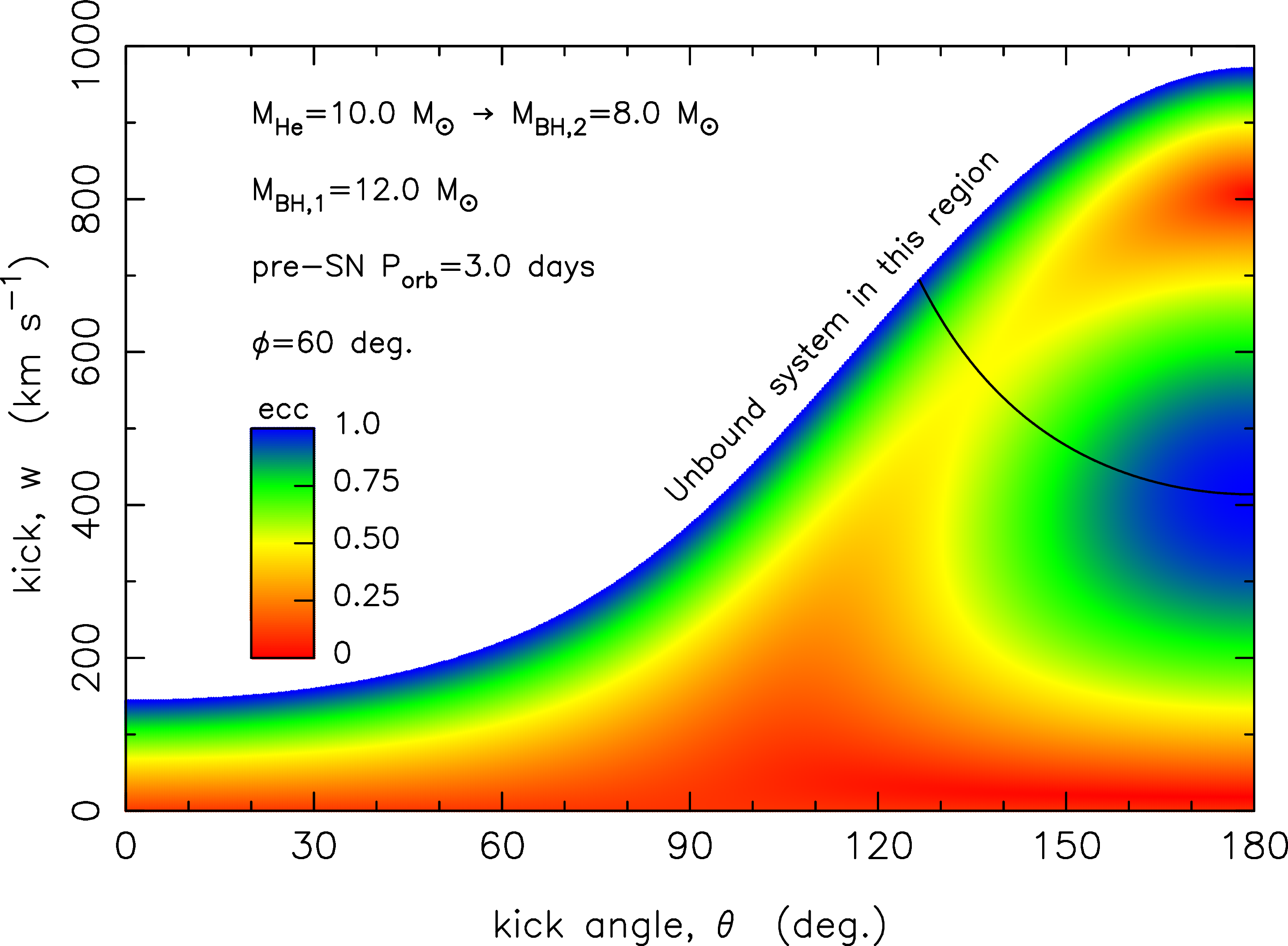} 
\includegraphics[width=0.50\textwidth]{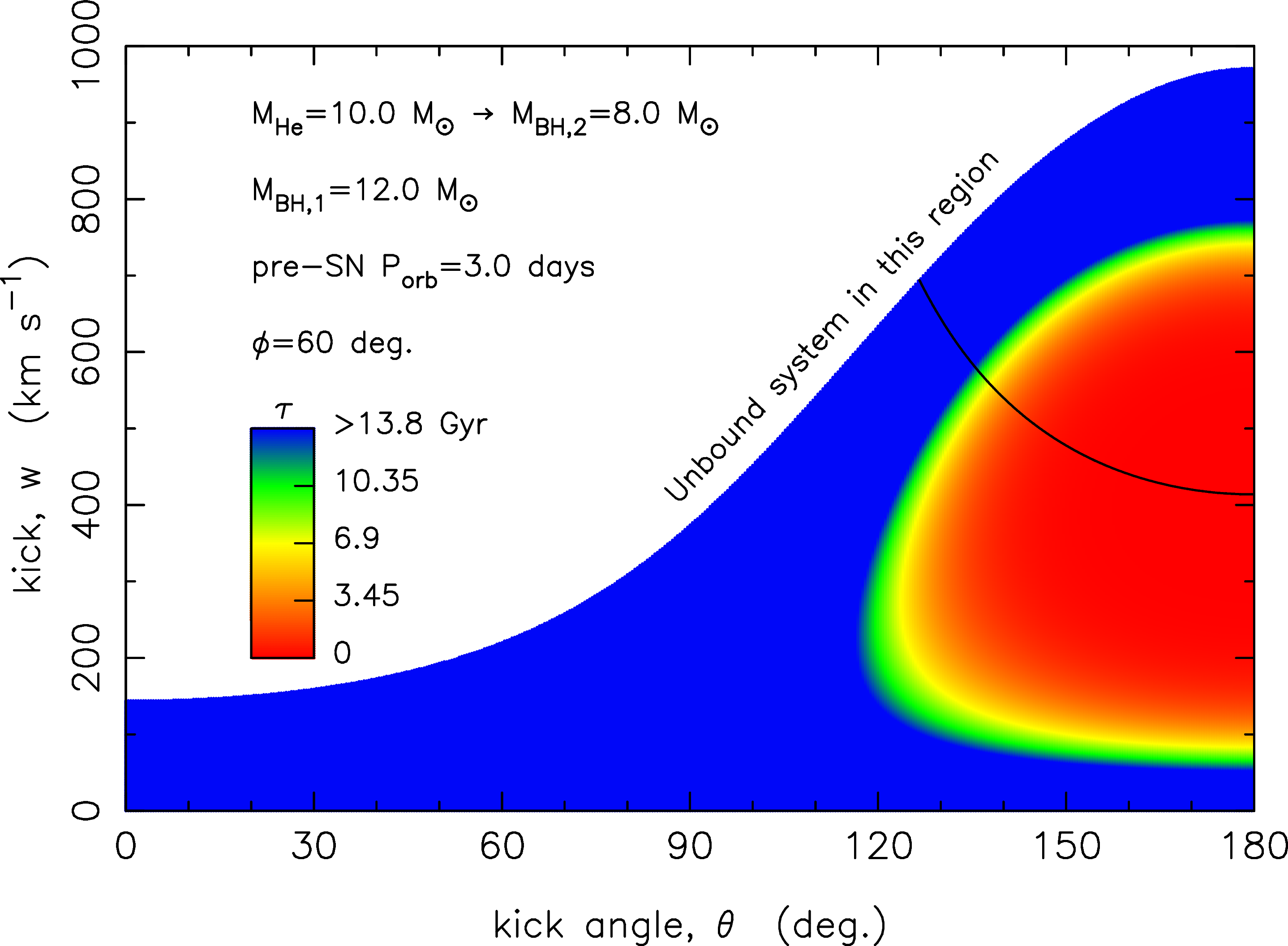} 
\caption{Post-SN parameters calculated from SN conditions similar to those of Figs.~\ref{fig:misalignment_kickangles} and \ref{fig:misalignment_theta-phi}, except that in all cases here the second kick angle, $\phi =60^\circ$ is fixed and the kick velocity, $w$ is a variable. Color coded (red to blue): misalignment angle, $\delta$ (top left); orbital period, $P_{\rm orb}$ (top right); eccentricity, $e$ (bottom left); and merger time,  $\tau_{\rm GW}$ (bottom right), as a function of kick angle, $\theta$ and kick velocity, $w$. 
The systems above the black line have a first-born BH with a retrograde spin because of the kick tilting the orbit, $\delta >90^\circ$.
\label{fig:param_kick-theta}}
\end{figure*}

\begin{figure*}
\vspace{-0.7cm}
\includegraphics[width=0.50\textwidth]{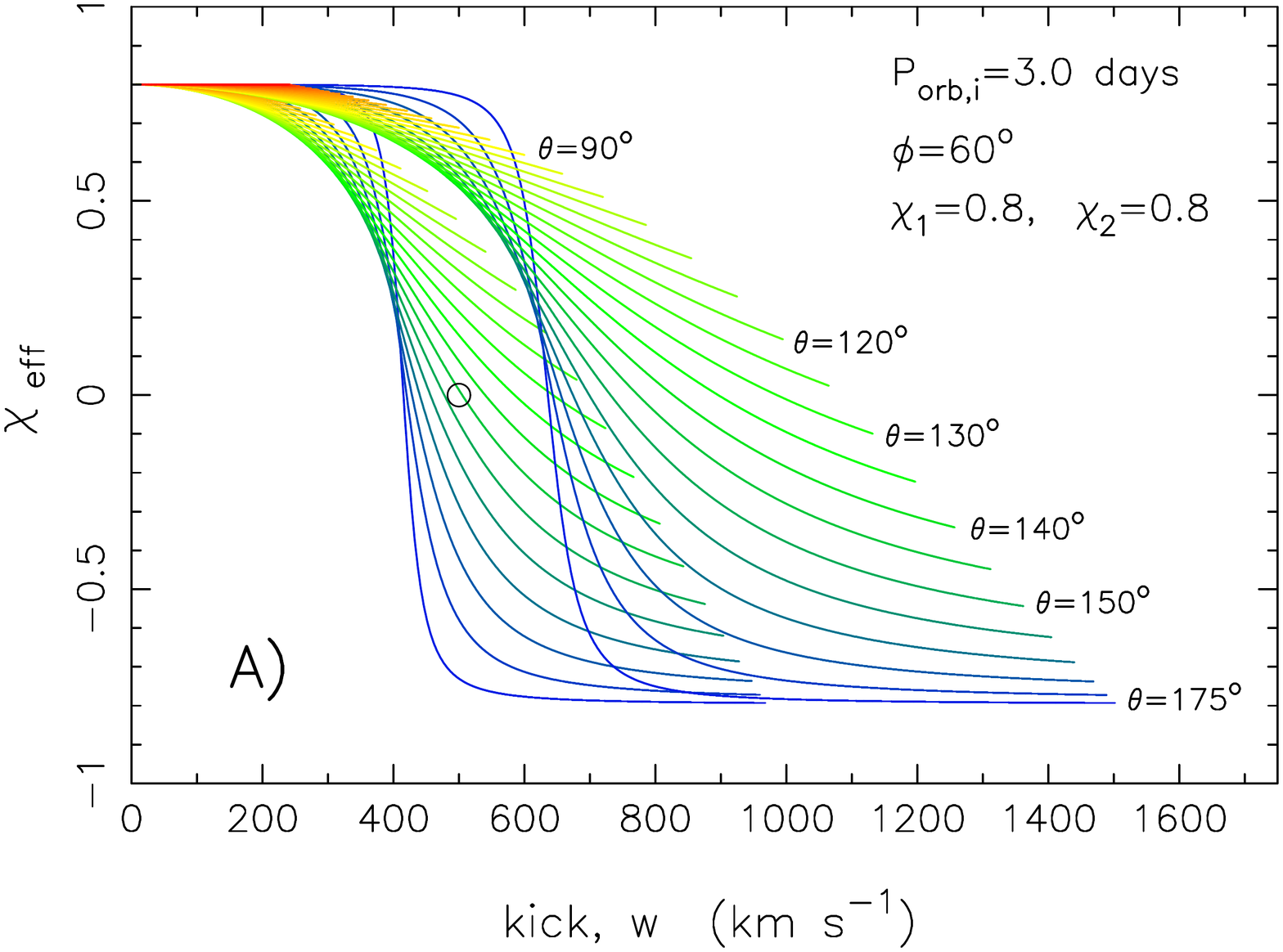}
\hspace{-1.0cm}
\includegraphics[width=0.50\textwidth]{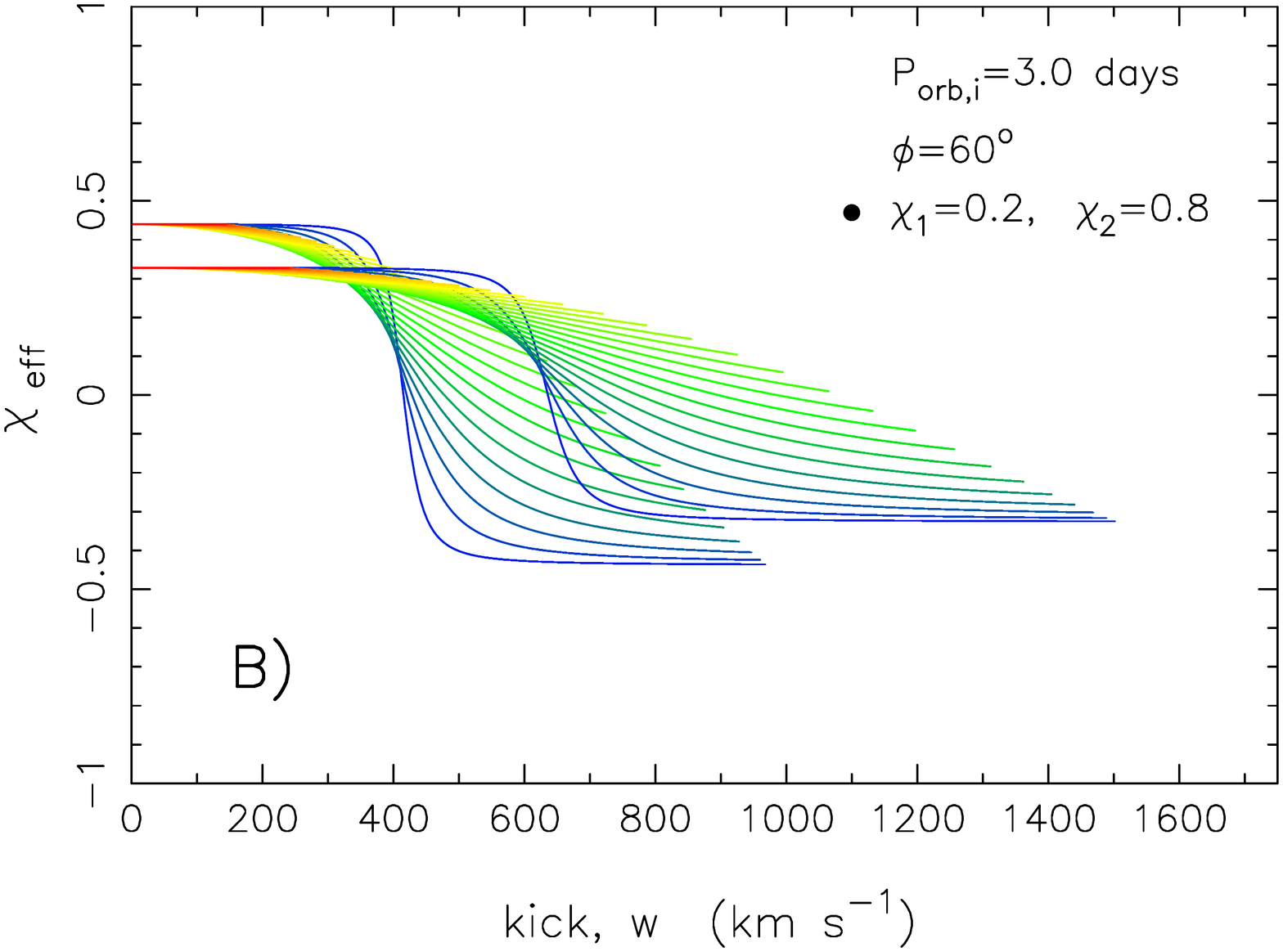}
\includegraphics[width=0.50\textwidth]{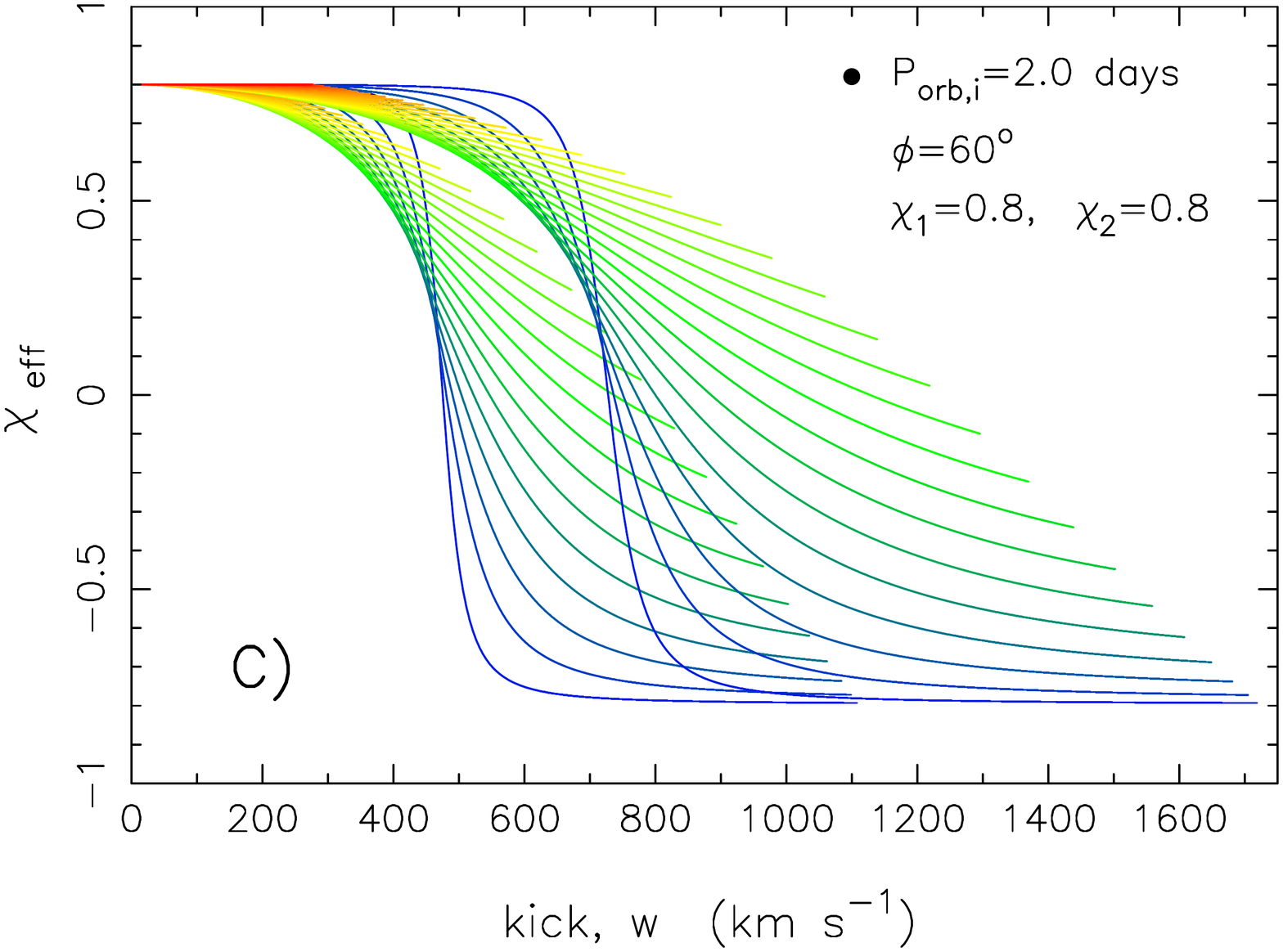}
\includegraphics[width=0.50\textwidth]{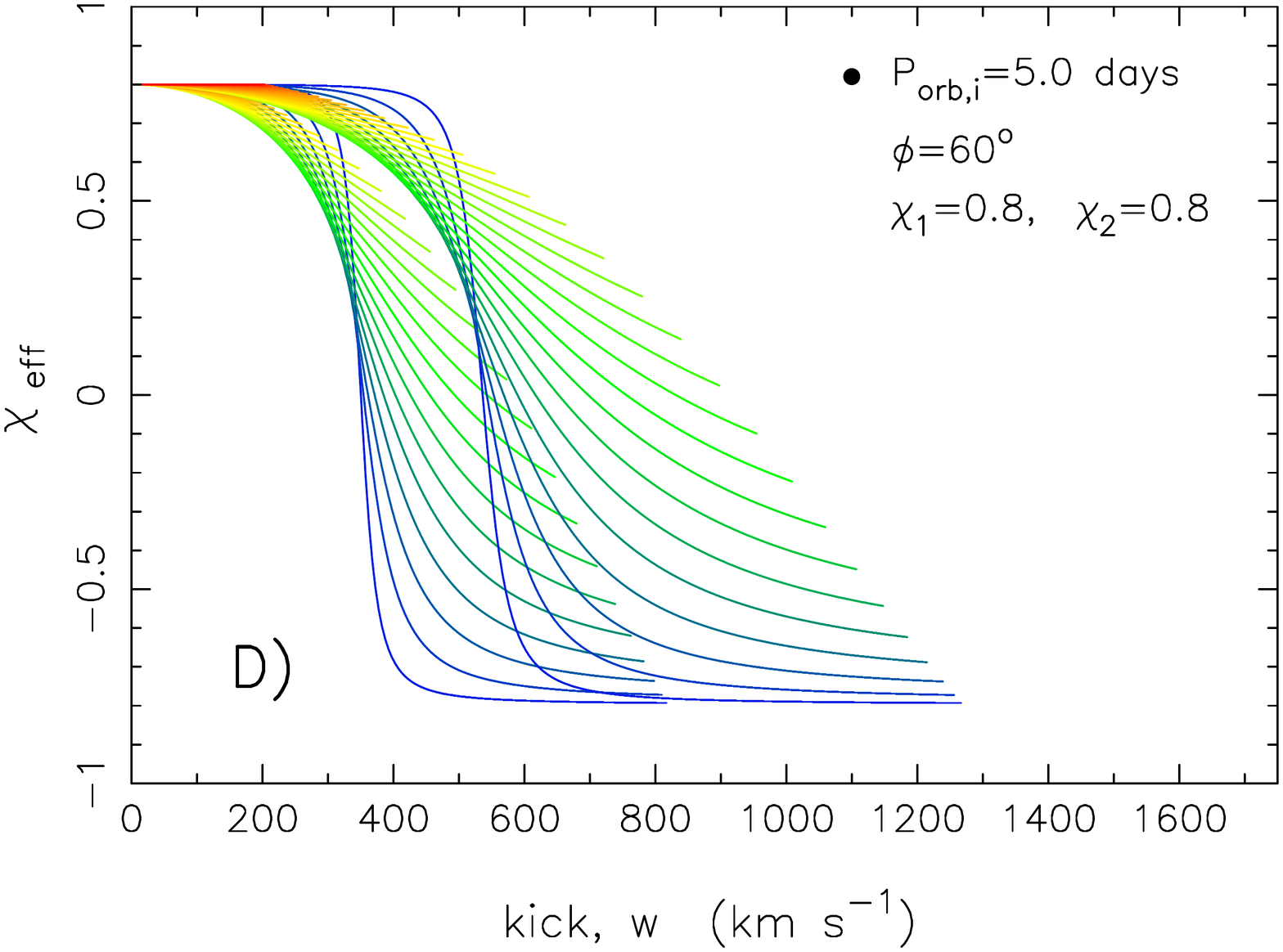}
\caption{Effective spin, $\chi_{\rm eff}$ as a function of kick velocity, $w$ and kick angle, $\theta$ (color-coded curves, red to blue with increasing $\theta$) imparted on the second-born BH for the same $12+8\;M_\odot$ BH+BH system as in Figs.~\ref{fig:misalignment_kickangles}--\ref{fig:param_kick-theta}. 
(The family of curves shifted to larger $w$ values are for $59+16\;M_\odot$ BH+BH systems.) Panels~A--D are for different pre-SN orbital periods, $P_{\rm orb,i}$ and BH component spins, $\chi_{1,2}$. The little open black circle in panel~A indicates $(w,\,\chi_{\rm eff})=(500\;{\rm km\,s}^{-1},\,0.0)$ where $\theta=\theta_{\rm retro}=145.8^{\circ}$. 
\label{fig:kick_chi_eff}}
\end{figure*}

Figures~\ref{fig:misalignment_kickangles} and \ref{fig:misalignment_theta-phi} show the misalignment angle, $\delta$ as a function of the two kick angles ($\theta$ and $\phi$) resulting from the production of a $12+8\;M_\odot$ BH+BH system, originating from a progenitor $12+10\;M_\odot$ BH+He-star binary with a fixed pre-SN orbital period of $P_{\rm orb}=3.0\;{\rm days}$ and applying a fixed kick velocity of $w=500\;{\rm km\,s}^{-1}$. The locations of the two critical angles, $\theta_{\rm crit}=105.4^{\circ}$ and $\theta_{\rm retro}=145.8^{\circ}$ are clearly seen in both figures.

Figure~\ref{fig:param_kick-theta} shows a variety of post-SN parameters calculated from the same progenitor system as in Figs.~\ref{fig:misalignment_kickangles} and \ref{fig:misalignment_theta-phi}, except that here the second kick angle is kept constant, $\phi =60^\circ$, and the kick velocity, $w$ is variable. The color coding in each panel refers to either: misalignment angle (top left), orbital period (top right), eccentricity (bottom left), or merger time (bottom right). 
The systems above the black curved line have a first-born BH with a retrograde spin because the kick tilts the orbit to $\delta >90^\circ$. 

Figure~\ref{fig:kick_chi_eff} shows examples of effective spin, $\chi_{\rm eff}$ as a function of kick velocity, $w$ and kick angle, $\theta$ (color-coded curves from red to blue with increasing value of $\theta$) imparted on the second-born BH for the same resulting $12+8\;M_\odot$ BH+BH system as in Figs.~\ref{fig:misalignment_kickangles}--\ref{fig:param_kick-theta}. 
The four different panels are calculated for different pre-SN orbital periods, $P_{\rm orb,i}$ and assumptions of fixed BH component spins, $\chi_{1,2}$. 

\subsubsection{Negative effective spins from large kicks}\label{subsubsec:negative_spins}
An important message from Figure~\ref{fig:kick_chi_eff} is that producing tight systems with $\chi_{\rm eff}<0$ that become BH+BH mergers within a Hubble time requires a kick of $w>v_{\rm rel}$, often corresponding to $w> 400\sim 600\;{\rm km\,s}^{-1}$. (Notice, that here retrograde systems are equivalent to $\chi_{\rm eff}<0$ because $\Theta_2\simeq \Theta_1=\delta$, see Section~\ref{subsubsec:spin-directions}.) It can thus be concluded that if BH spin-axis tossing is {\em not} at work then, to produce BH+BH mergers with $\chi_{\rm eff}<0$ in isolated systems, either very large secondary BH kicks are needed or pre-SN BH and stellar spins are not aligned --- two conditions which are both highly questionable (Section~\ref{subsec:BH_component_spins}).

To better demonstrate the effects on post-SN orbits and misalignment angles, $\Theta_1=\delta$ of the first-born BHs due to kicks, Figs.~\ref{fig:misalignment_kickangles}--\ref{fig:kick_chi_eff} display the results based on a relatively low-mass BH+BH system ($M_{\rm BH,1}=12.0\;M_\odot$, $M_{\rm He}=10.0\;M_\odot$ and $M_{\rm BH,2}=8.0\;M_\odot$). Even in that case, it is seen that a minimum kick velocity of $w>400\;{\rm km\,s}^{-1}$ (in a backwards direction, $\theta \rightarrow 180^\circ$) is needed to obtain a retrograde post-SN orbit (i.e. $\Theta_1=\delta > 90^\circ$ and thus $\chi_1 \cos\Theta_1<0$ and $\chi_{\rm eff}<0$). 

\subsubsection{Spin directions of progenitor systems}\label{subsubsec:spin-directions}
For isolated systems, the tilt angles of the two BH spins 
($\Theta_1$ and $\Theta_2$, see Eq.~\ref{eq:chi_eff2}) are assumed to be related to $\delta$ as follows. 
The tilt angle of the spin axis of the first-born BH ($\vec{\chi}_1$), by the time the system merges, is equal to the resulting misalignment angle, $\delta$ from the second SN in the BH+BH system, i.e. $\Theta_1=\delta$.  
This inference follows the simple argument \citep[e.g.][]{tv23,hut81,hil83,bv91,gt14,ba21} that at the time of the second SN, the first-born BH is assumed to possess prograde spin with its spin vector ($\vec{\chi}_1$) parallel to the pre-SN orbital angular momentum vector as a result of previous mass-transfer episodes from the progenitor of the collapsing stripped helium star onto the first-born BH (Section~\ref{subsec:BH_component_spins}).

For the spin tilt angle of the second-born BH, we consider two cases: 
In the case of {\em no tossing}: $\Theta_2\simeq \Theta_1=\delta$. This follows from the assumption that tidal torques will align the spin axis of the collapsing helium star with that of the orbital angular momentum vector prior to the SN. However, if the secondary BH is subject to {\em tossing} of its spin axis during its birth in the SN after core collapse, then $\Theta_2$ will follow a range of different (random) values (Section~\ref{subsubsec:BH_toss_PDF}).

\section{Monte Carlo simulations}\label{sec:simulations}
\subsection{Method and modelling}
After introducing the kinematic effects of the second SN on a sample system in Section~\ref{sec:SN_kinematics}, we now investigate the outcome on a large BH+BH population as function of various input distributions using Monte Carlo simulations. 
The assumptions about the input PDFs are kept relatively simple to better capture the effect of spin-axis tossing --- the main scope of this paper. The outlined method can readily be applied to input PDFs obtained from population synthesis. 

Here again, the starting point is a population of pre-SN binaries composed of the first-born BH and a companion star (the progenitor of the second-born BH) on the verge of core collapse. In the following, we first discuss the input distributions (PDFs) and refer to sample figures showing the outcome of the simulations along the way. A more systematic description of the displayed results is given in Section~\ref{sec:results} and discussed further in Section~\ref{sec:discussions}.

\subsubsection{Initial masses}
The masses of the first-born BH and the collapsing helium star companion were either chosen at fixed values or from a range of masses recovering most of the BH+BH events in the GWTC data \citep{aaa+22b}. 
In the majority of simulations demonstrated here (most panels in Fig.~\ref{fig:panel6} and Figs.~\ref{fig:panel6-3}--\ref{fig:panel6-5}), the mass component values were simply chosen from a flat PDF such that the resulting BH masses are in the range $M_{\rm BH,1}=8-70\;M_\odot$ and $M_{\rm BH,2}=3-45\;M_\odot$, thereby reproducing the bulk of detected BH+BH mergers (see Fig.~\ref{fig:GWTC3_BHBH} for GWTC data). 
The distribution of masses of the secondary (collapsing) stars are thus chosen in the range between $M_{\rm He}=3.75-56.25\;M_\odot$ because it is assumed here that 20\,\% of the mass of the collapsing star is lost during its collapse to a BH. This fraction may be on the high side. However, as the results are only weakly dependent on the BH mass distribution, using a mass-loss fraction of e.g. 5\,\% yields very similar results. For the same reason, the exact shape of the overall BH mass PDF is not very important here and it justifies the simple use of a flat PDF for the BH masses.

In the case of investigating BH spin-axis tossing effects for fixed BH masses, default pre-SN values of $37\;M_\odot$ and $30\;M_\odot$ are chosen for the first-born BH and the collapsing helium star, respectively (e.g. panels [A,\,B,\,E] in Fig.~\ref{fig:panel6-2}), to produce final BH+BH mergers with mass components of $M_{\rm BH,1}=37\;M_\odot$ and $M_{\rm BH,2}=24\;M_\odot$. 
Simulations with other fixed BH mass values, including more extreme mass ratios, are also shown in Fig.~\ref{fig:panel6-2}.

Finally, simulations were also performed assuming that some (or {\em all}) of the BH+BH systems have their mass components inverted (i.e. $M_{\rm BH,1}=24\;M_\odot$ and $M_{\rm BH,2}=37\;M_\odot$, panel~F in Fig.~\ref{fig:panel6-2}), to reflect an extreme extrapolation of a subpopulation of BH+BH progenitor systems undergoing mass reversal during their evolution --- such that the initially least massive (i.e. secondary star) of the two ZAMS stars accretes sufficient mass to end up producing the first-born BH; and the initial primary star then produces the second-born BH. For the general process of mass reversal and SN order in close binaries, see e.g. \citet{wl99,ts00,zdi+17}.

\subsubsection{Initial orbital separation}
The pre-SN orbital separation, $a_0$ is chosen from a flat PDF between: $4<a_0<40\;R_\odot$. The lower limit is set to fit the size of the collapsing helium star within its Roche lobe, and the upper limit is chosen such that the majority of the surviving post-SN systems will remain in orbits tight enough to let the BH+BH system merge within a Hubble time ($\sim 13.8\;{\rm Gyr}$). As demonstrated in the top panels of Fig.~\ref{fig:panel6-2}, the value of $a_0$ is not important for the distribution of $\chi_{\rm eff}$ (unless BH kicks are extreme {\em and} pre-SN orbits are very wide) --- only the number of systems merging within a Hubble time decreases with increasing $a_0$ (compare panels~A and B).

\subsubsection{BH kick}\label{subsubsec:BH-kick}
The kick imparted onto the newborn BH is assumed to have a random (isotropic) direction with $\theta \in [0;\,180^\circ]$ and $\phi \in [0;\,360^\circ]$. This is obtained with a PDF of $P(\theta)=0.5\sin(\theta)$ and a flat PDF for $\phi$. In most cases a kick magnitude of $w=50\;{\rm km\,s}^{-1}$ is applied. This modest kick value is chosen to match constraints from BHs in Galactic X-ray binaries \citep{man16,mir17} and also because the main emphasis here is to demonstrate the effect of the tossing of the BH spin axis. For comparison, however, simulations were also performed using $w=500\;{\rm km\,s}^{-1}$, and in a few cases without any kick ($w=0$, top row in Fig~\ref{fig:panel6-5}). It is found, perhaps somewhat surprising, that the distribution of $\chi_{\rm eff}$ does not depend much on $w$ (e.g. in Fig.~\ref{fig:panel6}, compare panels~C and D with panels~E and F). The main effect is simply that larger SN kicks destroy more binaries. For this reason, and to avoid too many free parameters, the default value is set to $w=50\;{\rm km\,s}^{-1}$. 
The reason for the (almost) kick invariance is that for a large part of the BH+BH progenitor systems the applied kick is still relatively small compared to the pre-SN orbital velocity ($w/v_{\rm rel}\ll 1$). 

\subsubsection{Spin magnitudes} 
\begin{figure}
\vspace{-0.5cm}\hspace{-0.6cm}
\includegraphics[width=0.55\textwidth]{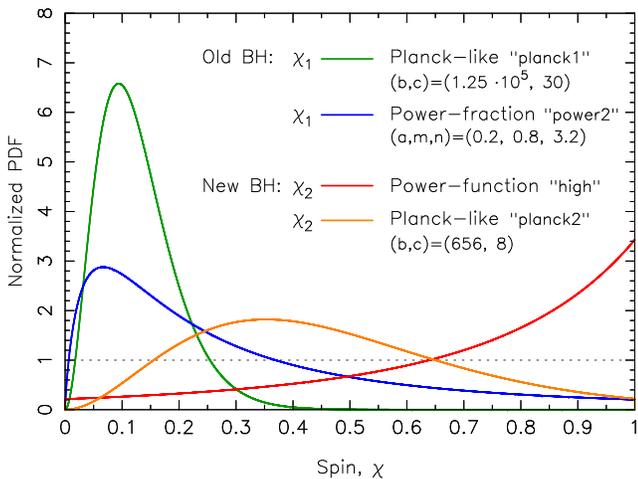}
\caption{Spin magnitude probability distribution functions (PDFs) of the first- ($\chi_1$) and second-born ($\chi_2$) BHs in isolated binary systems. The green, orange and blue curves reflect Planck-like PDFs: $P(\chi_1)=b\chi_1^3/(\exp(c\chi_1)-1)$, and a power-fraction PDF: $P(\chi_1)=(\chi_1/a)^m/(1+\chi_1/a)^n$, respectively. The red curve is a simple power-function PDF given by: $P(\chi_2)=24/(7\,(2-\chi_2)^4)$. See legends and Table~\ref{table:chi} for values of $b$, $c$, $a$, $m$ and $n$. The PDFs for $\chi_1$ and $\chi_2$ reflect different expectations of slow and fast-spinning BHs, respectively, based on binary star evolution (see text). 
For comparison, the grey dotted line is a flat PDF.
\label{fig:chi_dist}}
\end{figure}

The spin magnitudes of the two BH components ($\chi_1$ and $\chi_2$) are drawn from four different PDFs plotted in Fig.~\ref{fig:chi_dist} (see details in Table~\ref{table:chi}), based on different expectation values for the first- and second-born BH in an isolated binary system (see Section~\ref{subsec:BH_component_spins} for additional discussions). Whereas the first-born BHs have been argued to be preferentially slowly spinning \citep[$\chi_1\lesssim 0.1$,][]{fm19,bfq+20}, the second-born BHs are expected to spin much faster due to tidal interactions between the first-born BHs and the helium stars that collapse and produce the second-born BHs \citep{kzkw16,hp17,zkk18}. 

The two distributions ``planck1'' and ``planck2'' have median $\chi$ values of $\chi_1\sim 0.1$ and $\chi_2\sim 0.4$, respectively, and were selected to mimic the empirical distributions inferred in fig.~17 in \citet{aaa+22b} based on GWTC data of BH+BH mergers. These are the distributions applied in the default results shown in Fig.~\ref{fig:panel6} [panels~A,\,B]. 
The effects on the $\chi_{\rm eff}$ distribution, however, were primarily investigated here using the alternative PDFs ``power2'' and ``high'' for $\chi_1$ and $\chi_2$, respectively (Fig.~\ref{fig:panel6} [panels~C--F], and Fig.~\ref{fig:panel6-2}), because these distributions likely better reflect expectations from close binary star evolution (Section~\ref{subsec:BH_component_spins}).  
Finally, the effects on $\chi_{\rm eff}$ by applying fixed or a narrow range of values of $\chi_1$ and/or $\chi_2$ were also investigated, see Figs.~\ref{fig:panel6-3}--\ref{fig:panel6-5}. 

\begin{table*}
\caption{Normalized PDFs of BH component spins, $\chi_i$ ($i=1,2$) for simulating $\chi_{\rm eff}$ of BH+BH systems that merge. See plots of the PDFs in Fig.~\ref{fig:chi_dist} and further discussions in Section~\ref{subsec:BH_component_spins}. \label{table:chi}}
\begin{center}
\begin{tabular}{lllcc}
\hline
name & PDF & shape & average & median\\
\hline
\noalign{\smallskip} 
planck1 &   $P(\chi_i)=\displaystyle\frac{125\,000\;\chi_i^3}{\exp(30\,\chi_i)-1}$ & strong peak near $\chi_i\sim 0.1$ & $\langle \chi_i \rangle =0.128$ & $\chi_{i,\rm med}=0.117$\\
\noalign{\smallskip} 
\noalign{\smallskip}  
power2  &   $P(\chi_i)=\displaystyle\frac{17.4\,(\chi_i/0.2)^{0.8}}{(1+\chi_i/0.2)^{3.2}}$  & concentrated at small $\chi_i$ & $\langle \chi_i \rangle =0.288$ & $\chi_{i,\rm med}=0.214$\\
\noalign{\smallskip}
\noalign{\smallskip} 
planck2 &   $P(\chi_i)=\displaystyle\frac{656\;\chi_i^3}{\exp(8\,\chi_i)-1}$  & broad peak $\chi_i\sim 0.2-0.6$ & $\langle \chi_i \rangle =0.450$ & $\chi_{i,\rm med}=0.426$\\
\noalign{\smallskip}  
\noalign{\smallskip}
high    &   $P(\chi_i)=\displaystyle\left(\frac{24}{7}\right)\,\frac{1}{(2-\chi_i)^4}$  & increasing with $\chi_i$ & $\langle \chi_i \rangle =0.714$ & $\chi_{i,\rm med}=0.789$\\ 
\noalign{\smallskip}  
\hline
\end{tabular}
\end{center} 
\end{table*}

\subsubsection{BH spin-axis tossing}\label{subsubsec:BH_toss_PDF} 
The PDF for the degree of tossing of the spin axis of the second-born BH is assumed to follow an isotropic distribution of post-SN spin axis directions: $P(\Theta_2)=0.5 \sin (\Theta_2)$, where $\Theta_2 \in [0;\,180^\circ]$. In the case of a core collapse without the effect of BH spin-axis tossing, it is assumed that: $\Theta_2=\Theta_1=\delta$ (see Section~\ref{subsubsec:spin-directions}).

\section{Results}\label{sec:results}
A number of histograms are presented in Figs.~\ref{fig:panel6}--\ref{fig:panel6-5} to display the resulting $\chi_{\rm eff}$ distributions of BH+BH systems produced from Monte Carlo simulations {\em with} (green bins) and {\em without} (red bins) BH tossing, following the methodology and details outlined in Sections~\ref{sec:SN_kinematics} and \ref{sec:simulations}.
The total number of simulated pre-SN systems is quoted in each panel of all figures. Underneath is stated the number of post-SN systems which survive and produce a BH+BH merger within a Hubble time (``GW systems''). A number of post-SN systems are disrupted (also stated in the legend) if the applied SN kick value is large ($w=500\;{\rm km\,s}^{-1}$). The rest of the surviving systems end up in BH+BH binaries that are too wide to merge within a Hubble time. (A very few systems $\mathcal{O}(<10^{-5})$ actually produce {\em direct} mergers if the second-born BH is kicked hard directly towards the first-born BH such that their closest spatial separation is less than the sum of their event horizon radii). 

Notice that the total number of systems simulated in each panel is simply adjusted to optimize the height of the most frequent bins in the histograms. Similarly, the plotted empirical GWTC-3 data bins have been multiplied with a constant scale factor of 100 for clarity. 
Some panels include solid curves which represent: (blue) the mean estimate of the PDF of $\chi_{\rm eff}$ from GWTC-3 data \citep{aaa+22b}, (grey) the 5\% and 95\% quantiles, and (black) the blue curve normalized to number of simulated systems. 
An example is shown in Fig.~\ref{fig:panel6}, panel~A, which may be considered the default result for BH spin-axis tossing. Note, the black curve is {\em not} a fit to the simulated distribution shown as the green histogram --- it is the normalized empirical data (but the agreement is remarkable).

The main result of the investigation in this work is that the observed distribution of $\chi_ {\rm eff}$ cannot be reproduced from isolated binaries {\em without} tossing of the BH spin axis (as clearly demonstrated in Fig.~\ref{fig:panel6} [panels~B,\,D,\,F], Fig.~\ref{fig:panel6-3}, and Fig.~\ref{fig:panel6-5} [panels~B,\,D,\,F]), but see also Section~(\ref{subsubsec:negative_spins}). 
On the other hand, simulations {\em including} tossing of the BH spin axis reproduce the observed distribution of $\chi_ {\rm eff}$ quite well --- even for a wide range of input parameter values. 
The resulting $\chi_{\rm eff}$ distribution is basically invariant to SN kicks, exact pre-SN orbital separations, and only weakly dependent on the input distributions of BH masses. This is demonstrated by comparing simulation results e.g. in Fig.~\ref{fig:panel6} [panels~C,\,E], Fig.~\ref{fig:panel6-2}, and Fig.~\ref{fig:panel6-5} [panel~A], as will be discussed further below in more detail. 
The few observed BH+BH systems with very large values of $\chi_{\rm eff}\sim 0.7$ are somewhat outliers and may be recovered using different PDFs for the BH spins (see Fig.~\ref{fig:panel6} [panel~C]), or these systems may simply have a distinct origin (Section~\ref{subsec:outliers}). 
We now discuss the results of each of the histograms.

Figure~\ref{fig:panel6} directly compares the resulting $\chi_{\rm eff}$ distribution {\em including} (left column) and {\em excluding} (right column) the process of BH spin-axis tossing. 
Here it is clearly demonstrated that only by including BH tossing the observed $\chi_{\rm eff}$ distribution can be reproduced.  
As an example, panel~A shows a simulation of the expected distribution of $\chi_{\rm eff}$ following a scenario with BH tossing. The plot is based on 25\,000 computations of the second SN, from which 22\,254 BH+BH systems are produced that will merge within a Hubble time. 
The agreement with the observed data (blue bins and black curve PDF) is striking --- see also Fig.~\ref{fig:panel6-5}, panel~A. 
Panel~C of Fig.~\ref{fig:panel6} is similar to panel~A, except that other spin distributions (``power2'' and ``high'', see Table~\ref{table:chi}) are applied to enable production of slightly larger $\chi_{\rm eff}$ values. Panels~B and D show similar simulations without BH tossing, which irrefutably cannot reproduce the observational data. The bottom row shows that the $\chi_{\rm eff}$ distribution is basically invariant to BH kick magnitude (Section~\ref{subsubsec:BH-kick}). Unless absurdly large, applying a kick of e.g. $w=500\;{\rm km\,s}^{-1}$ only increases the number of systems disrupted due to the SN. 

Figure~\ref{fig:panel6-2} assumes fixed BH masses. In panels~[A,\,B,\,E], the BH masses are equal to the average BH masses of $M_1=37\;M_\odot$ and $M_2=24\;M_\odot$ from GWTC-3, see Fig.~\ref{fig:GWTC3_BHBH}) and the simulations here show that the resulting $\chi_{\rm eff}$ distribution to a large degree is invariant to pre-SN orbital separation, $a_0$.
In panel~A, a constant pre-SN orbital separation value of $a_0=4\;R_\odot$ is applied, whereas $a_0=40\;R_\odot$ is applied in panel~B. Despite a factor of 10 in difference in the choice of $a_0$, the only major difference in the outcome is that fewer systems merge within a Hubble time if $a_0=40\;R_\odot$, as expected. (Notice that no post-SN systems will merge within a Hubble time if $a_0>54\;R_\odot$ and $w=50\;{\rm km\,s}^{-1}$.)
Panels~C--F show that the $\chi_{\rm eff}$ distribution depends only somewhat weakly on the exact choice of BH masses.

Figure~\ref{fig:panel6-3}, and panels~[B,\,D,\,F] of both Figs.~\ref{fig:panel6} and ~\ref{fig:panel6-5}, clearly demonstrate that reproducing the observed $\chi_{\rm eff}$ distribution without BH spin-axis tossing is impossible, no matter the choice of input distributions of BH masses, their spins and applied kicks (but see caveat in Section~\ref{subsubsec:negative_spins}). To boost the kinematic effect of SNe, in terms of producing small or negative values of $\chi_{\rm eff}$, very large kicks of $w=500\;{\rm km\,s}^{-1}$ were applied in all panels in Fig.~\ref{fig:panel6-3}. Despite of this, the simulated data which neglects BH spin-axis tossing cannot reproduce the observed data.

\subsection{Spin of first-born BH}
\begin{figure*}
\vspace*{-1.0cm}
\hspace*{-0.8cm}
\includegraphics[width=0.58\textwidth]{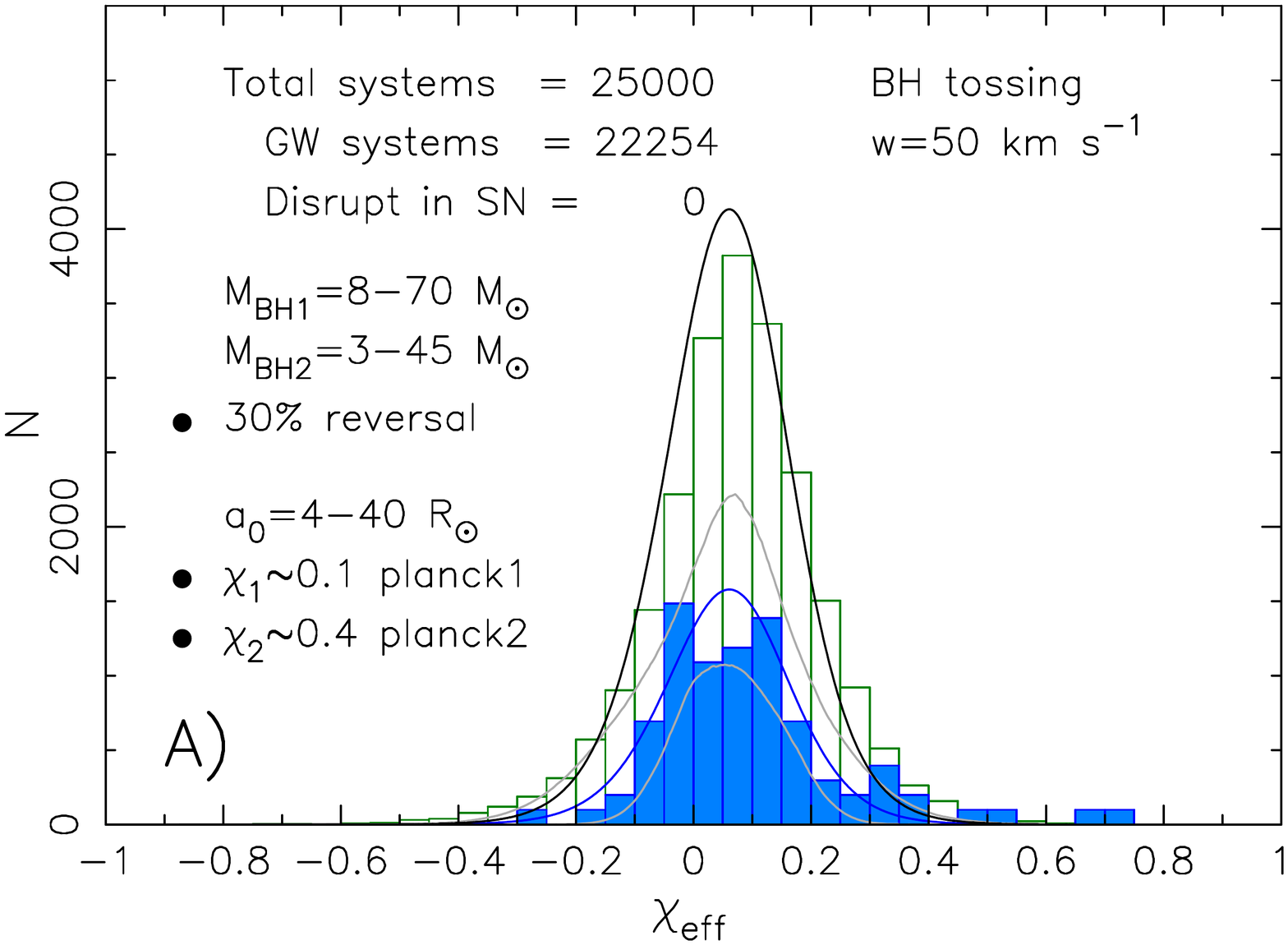}
\hspace*{-1.3cm}
\includegraphics[width=0.58\textwidth]{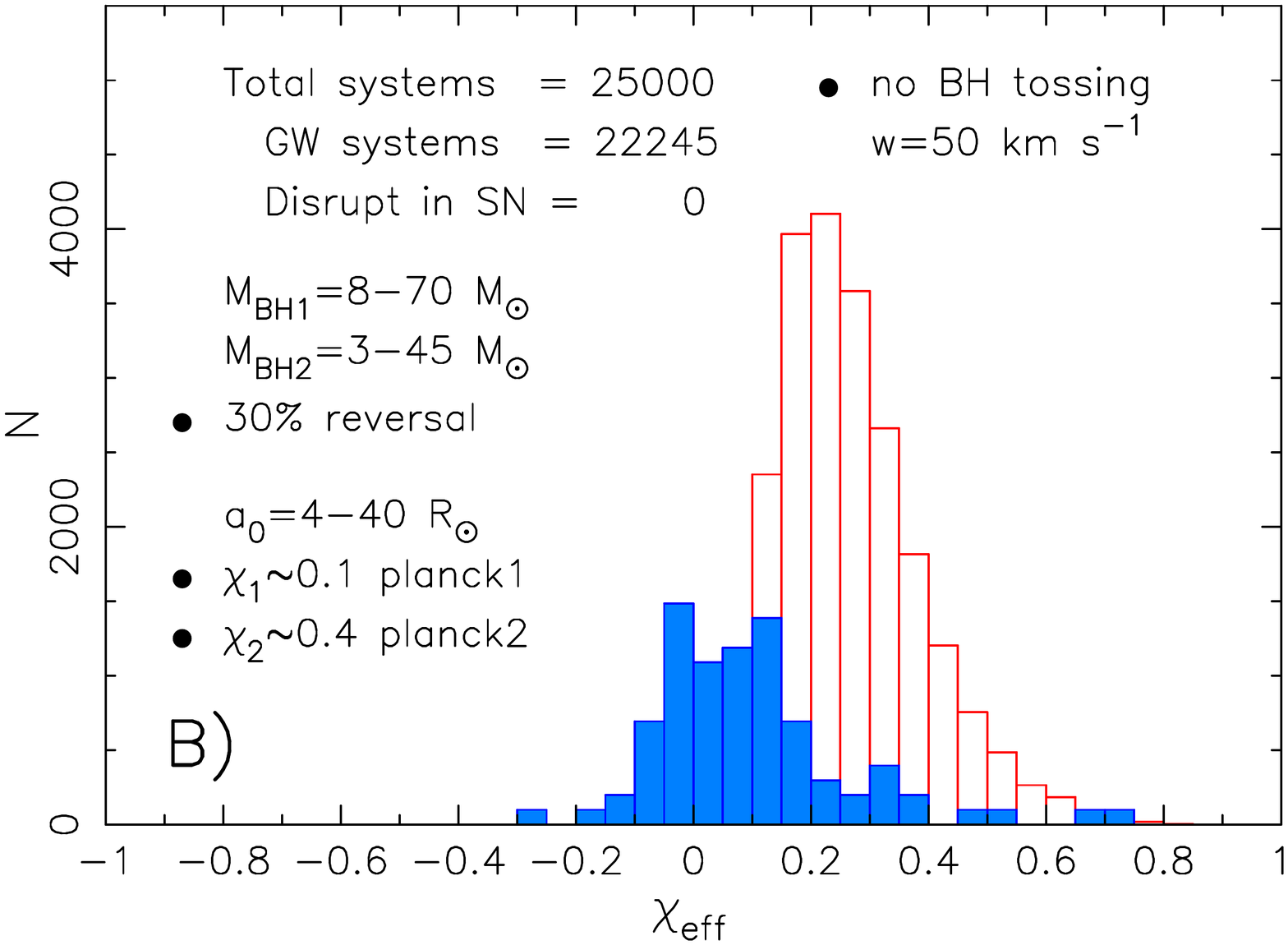}

\vspace*{-1.2cm}
\hspace*{-0.8cm}
\includegraphics[width=0.58\textwidth]{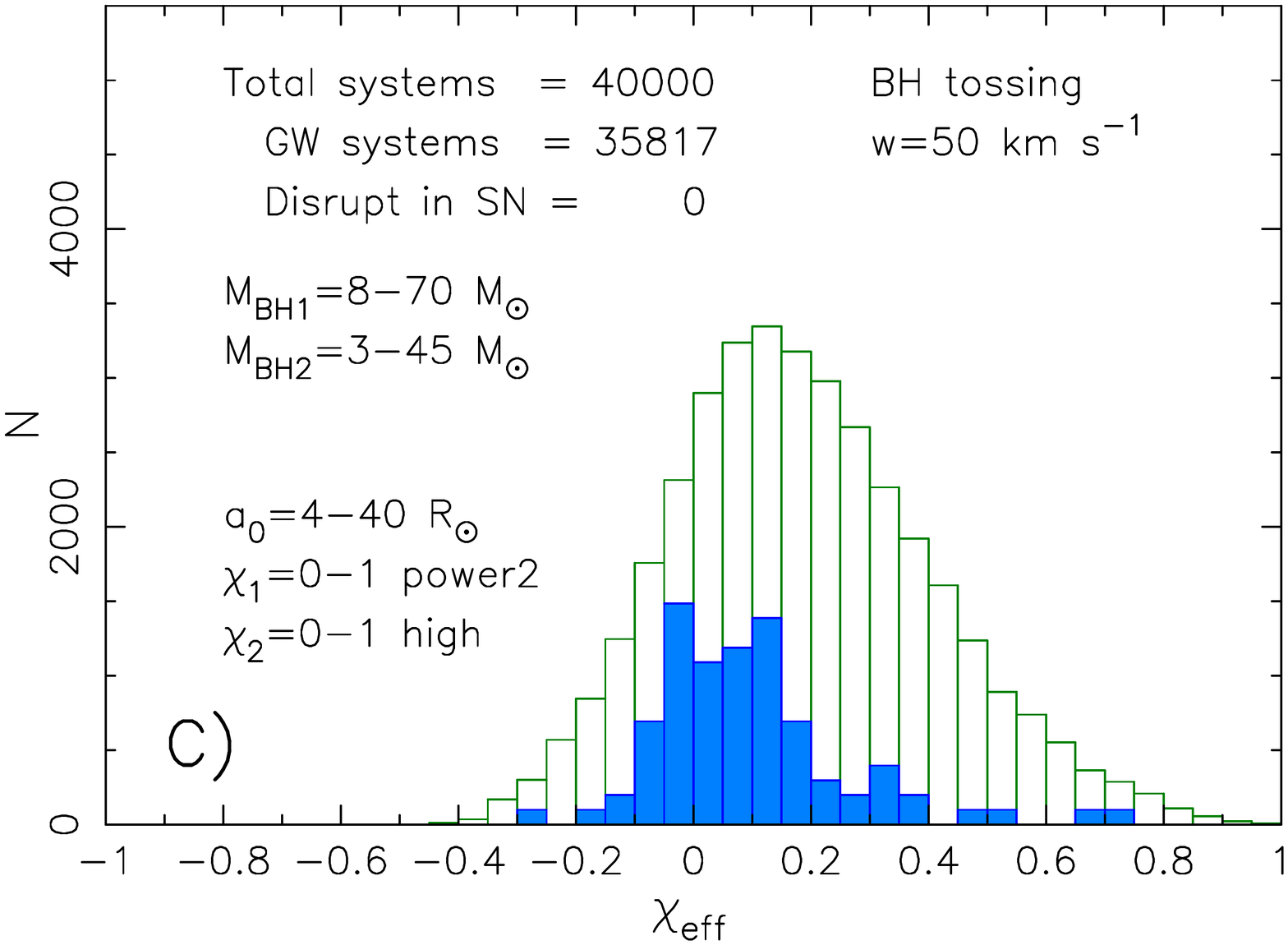}
\hspace*{-1.3cm}
\includegraphics[width=0.58\textwidth]{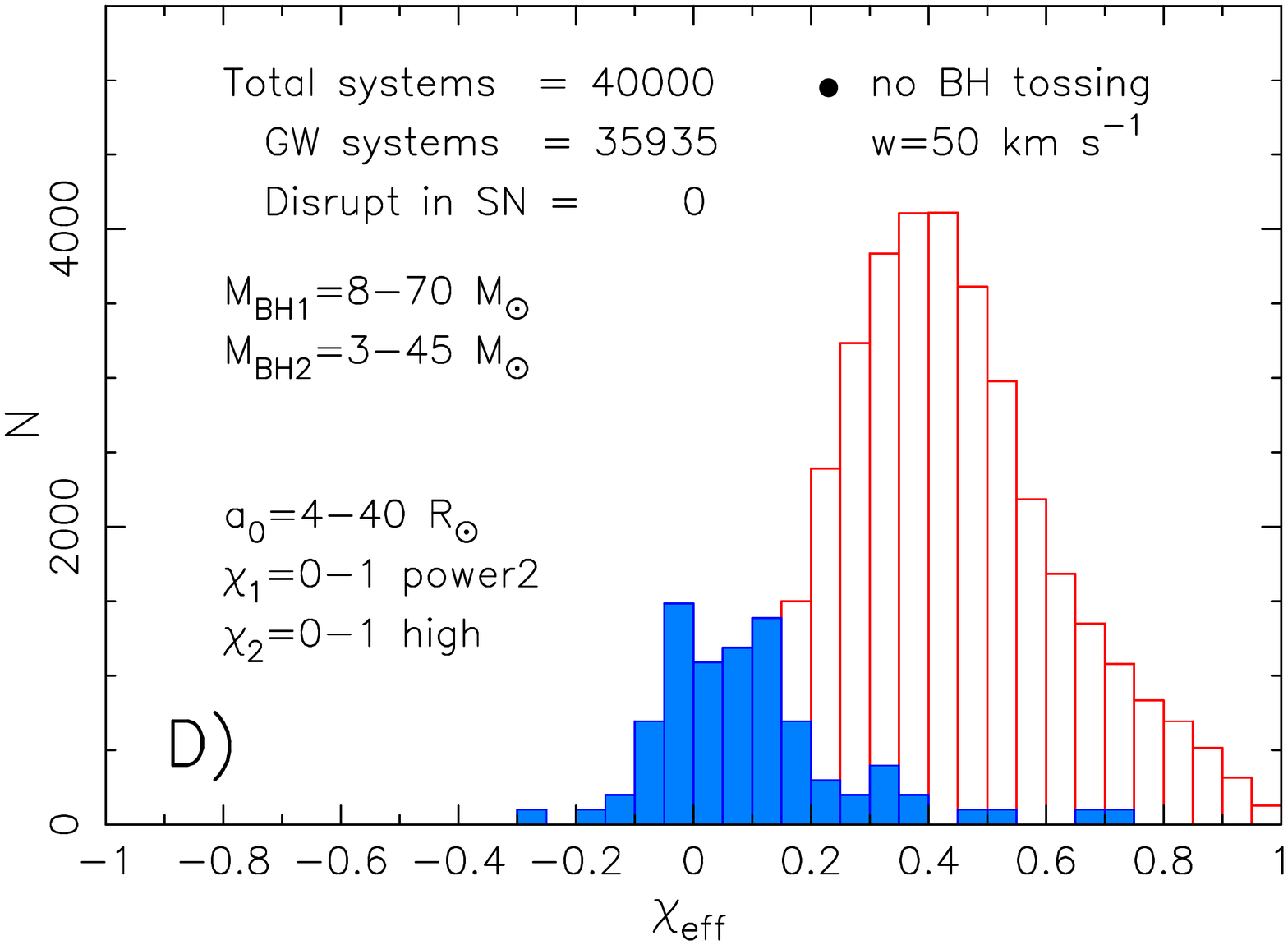}

\vspace*{-1.2cm}
\hspace*{-0.8cm}
\includegraphics[width=0.58\textwidth]{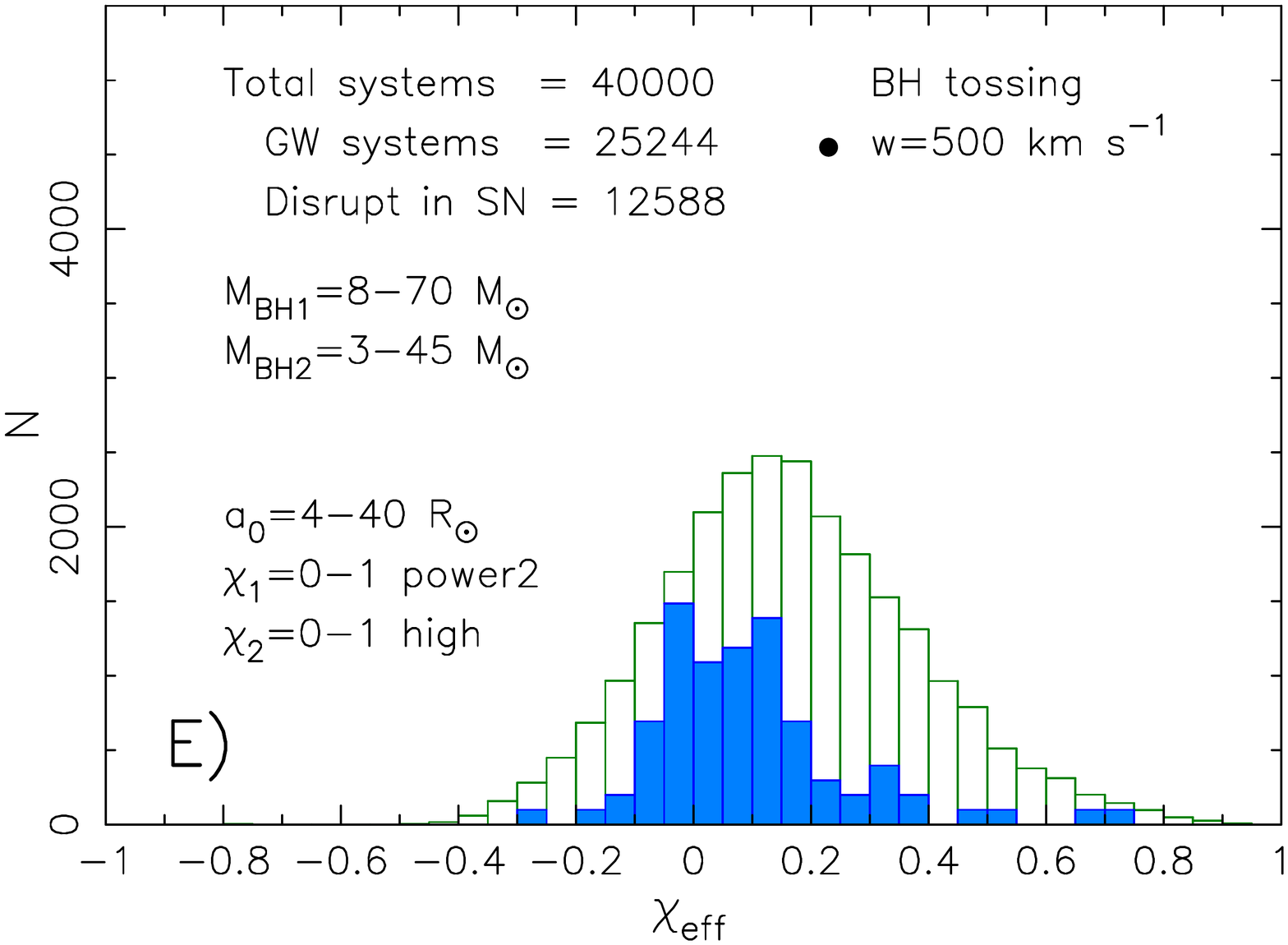}
\hspace*{-1.3cm}
\includegraphics[width=0.58\textwidth]{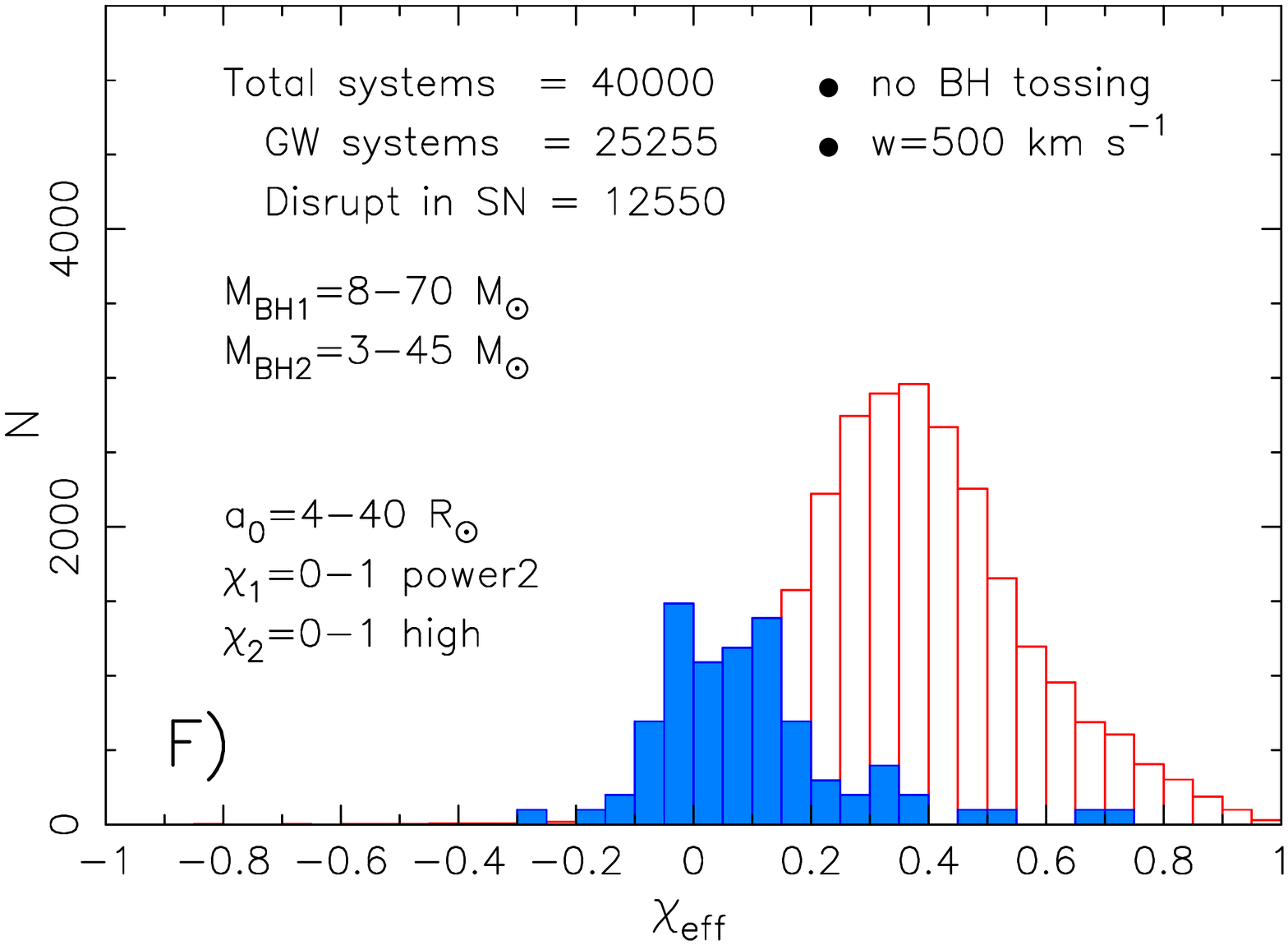}

\vspace*{-0.3cm}
\caption{Distributions of $\chi_{\rm eff}$ from observed GWTC-3 BH+BH data (solid blue bins) and simulated BH+BH systems (open green or red bins). The left-hand column (green bins) includes tossing of the second-born BH spin axis and leads to remarkable aggrement with data, whereas the right-hand column (red bins) assumes no BH tossing. Differences in assumed parameters for simulations in the top, middle and bottom rows are marked in the legend with a black bullet point and explained in the text.
\label{fig:panel6}}
\end{figure*}

\begin{figure*}
\vspace*{-1.0cm}
\hspace*{-0.8cm}
\includegraphics[width=0.58\textwidth]{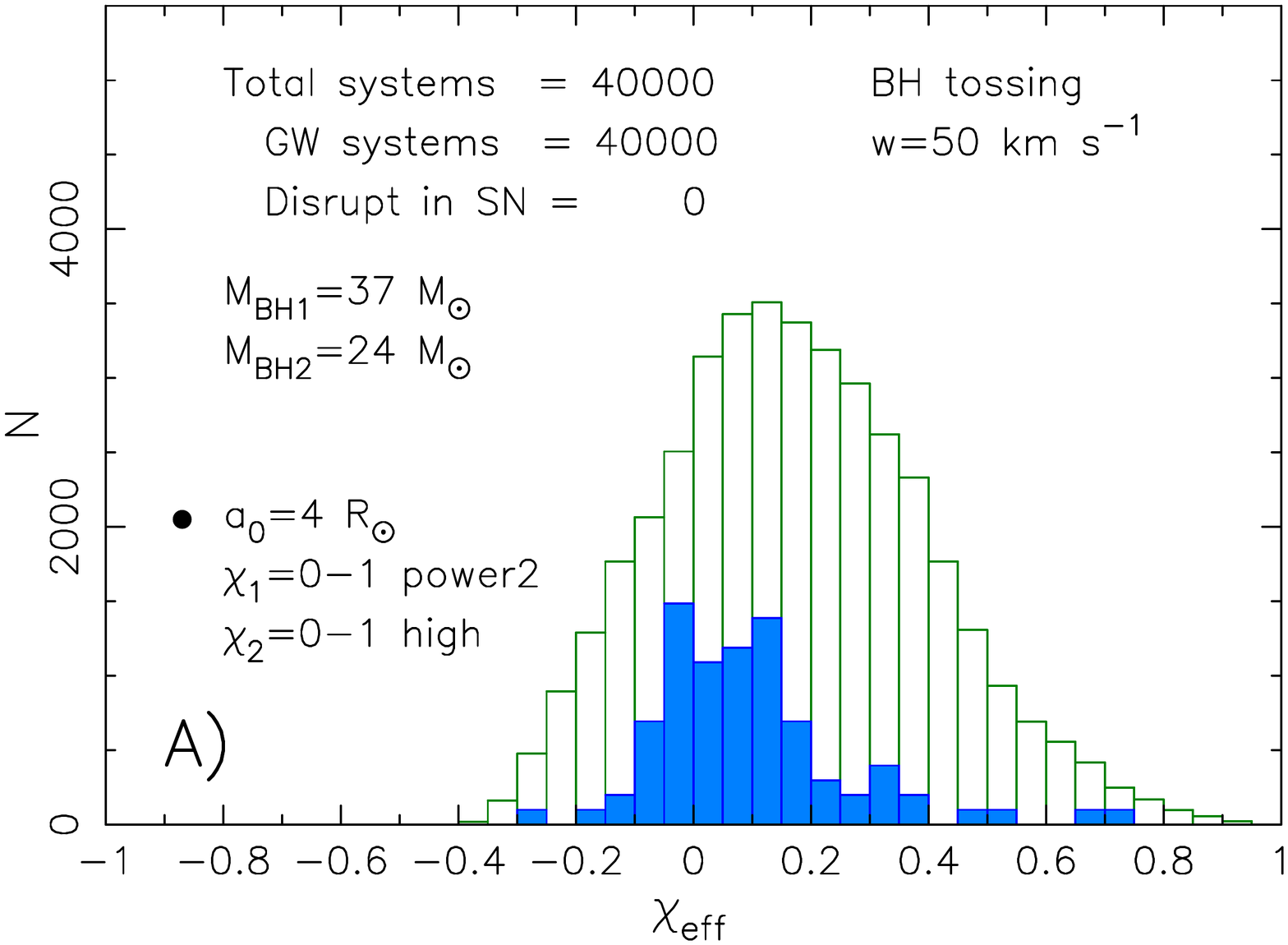}
\hspace*{-1.3cm}
\includegraphics[width=0.58\textwidth]{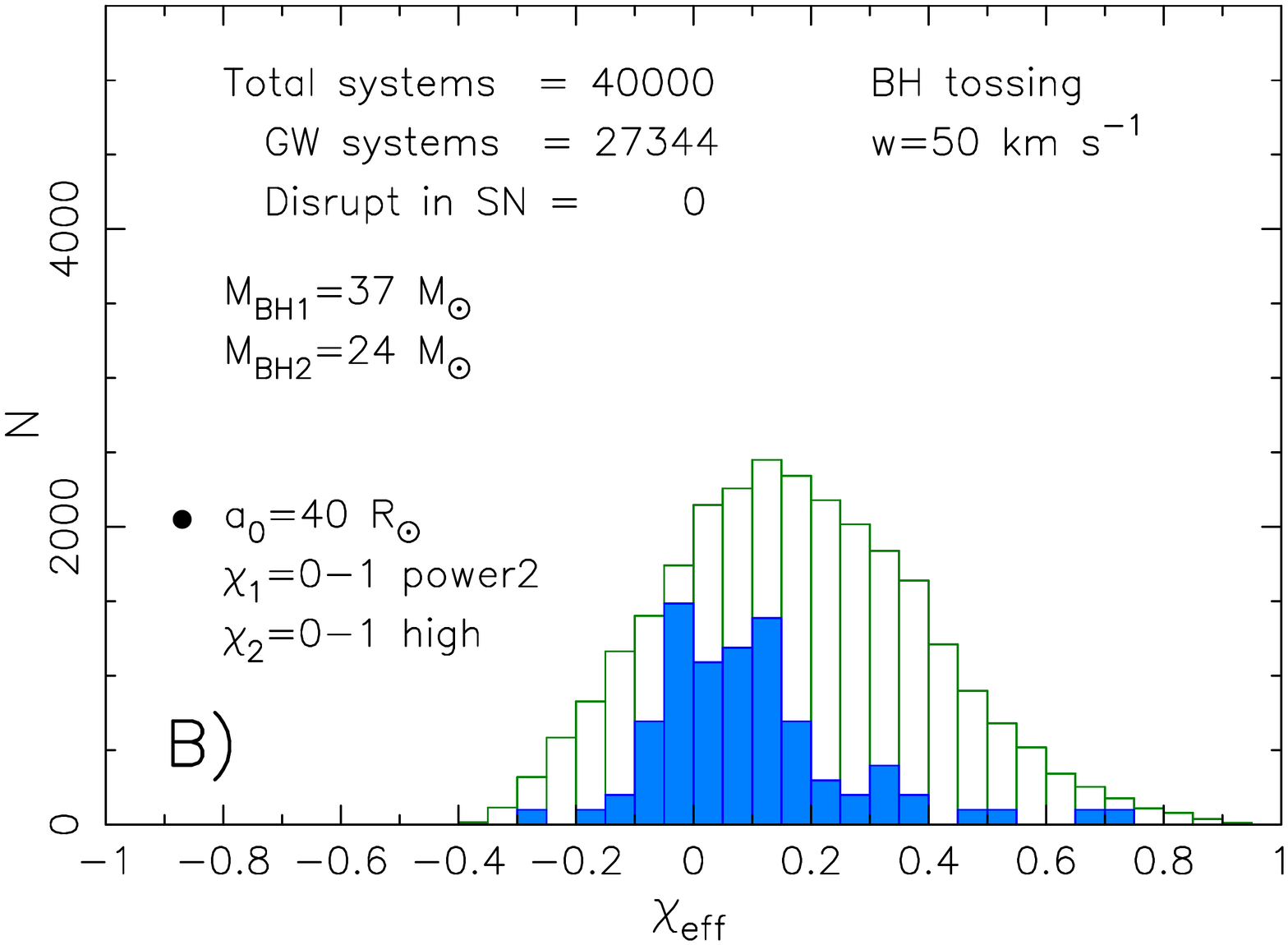}

\vspace*{-1.2cm}
\hspace*{-0.8cm}
\includegraphics[width=0.58\textwidth]{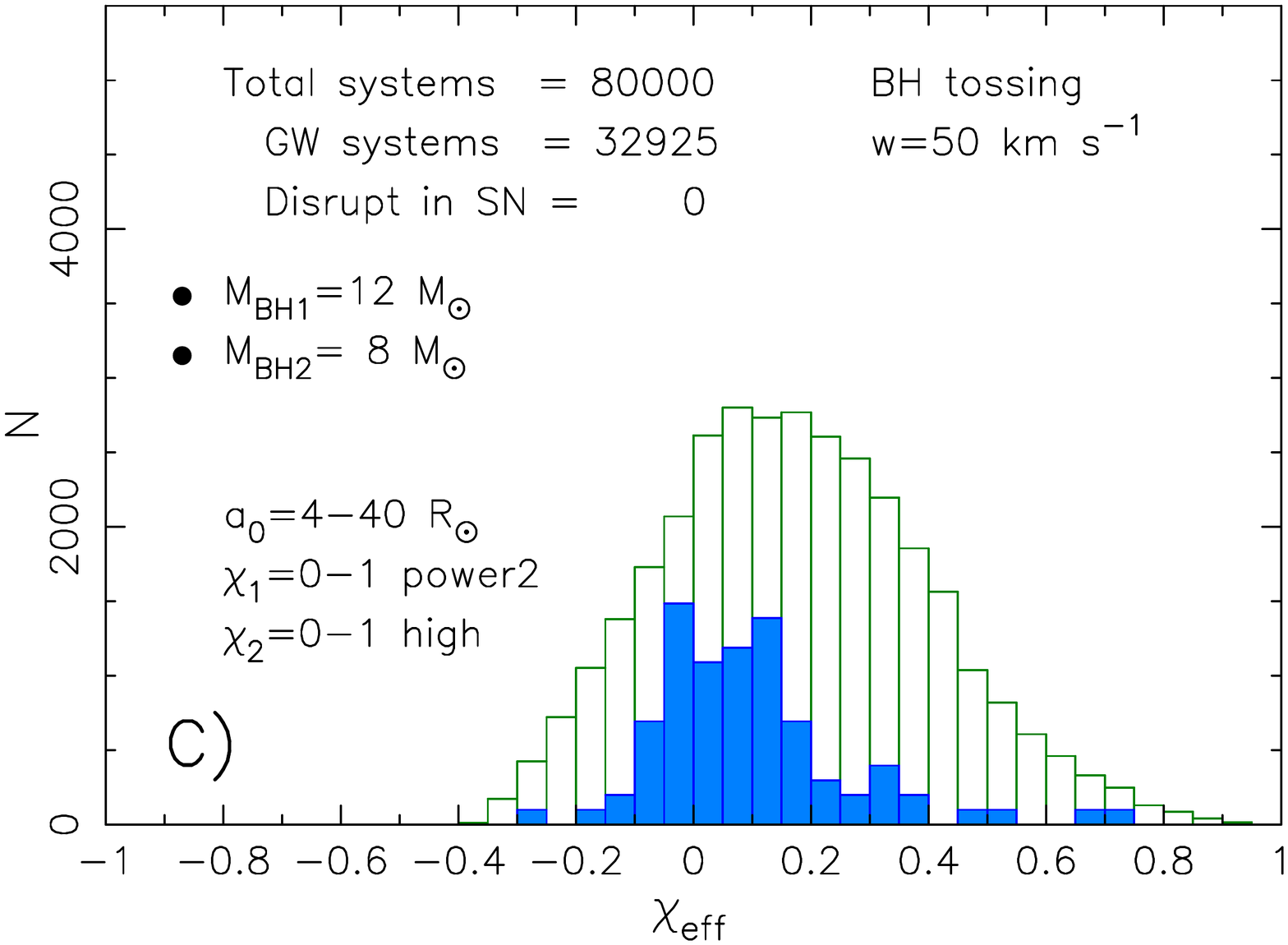}
\hspace*{-1.3cm}
\includegraphics[width=0.58\textwidth]{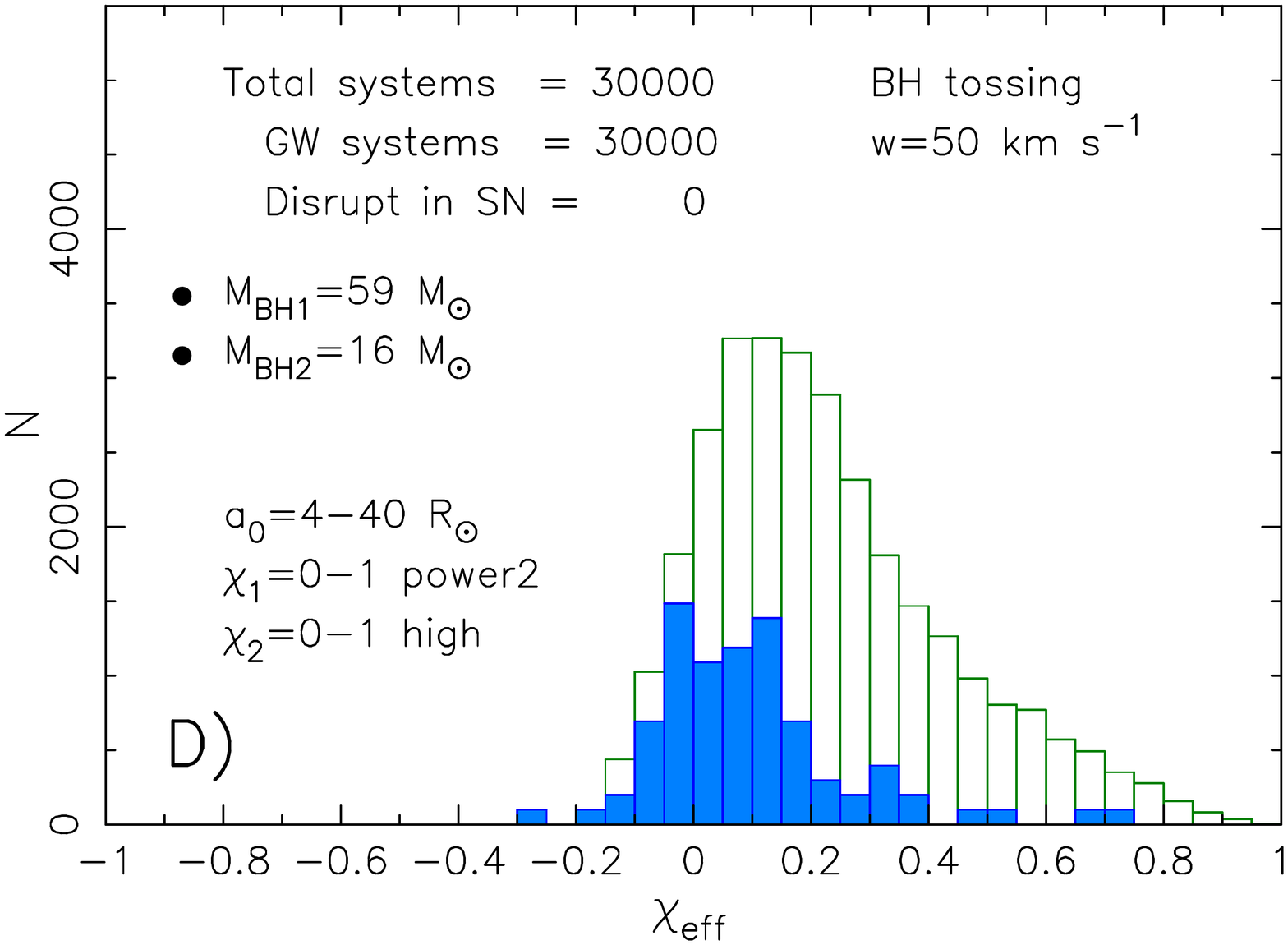}

\vspace*{-1.2cm}
\hspace*{-0.8cm}
\includegraphics[width=0.58\textwidth]{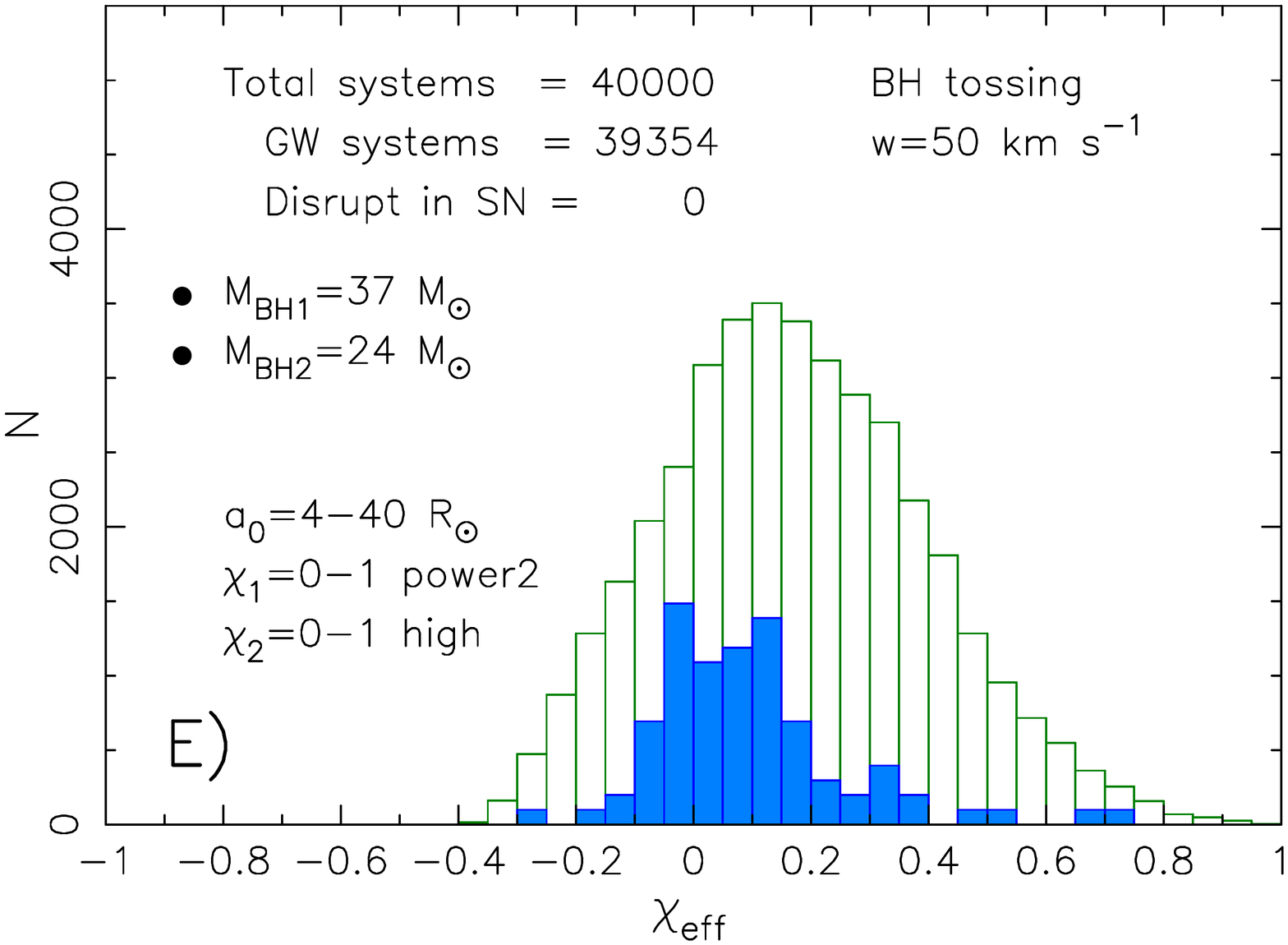}
\hspace*{-1.3cm}
\includegraphics[width=0.58\textwidth]{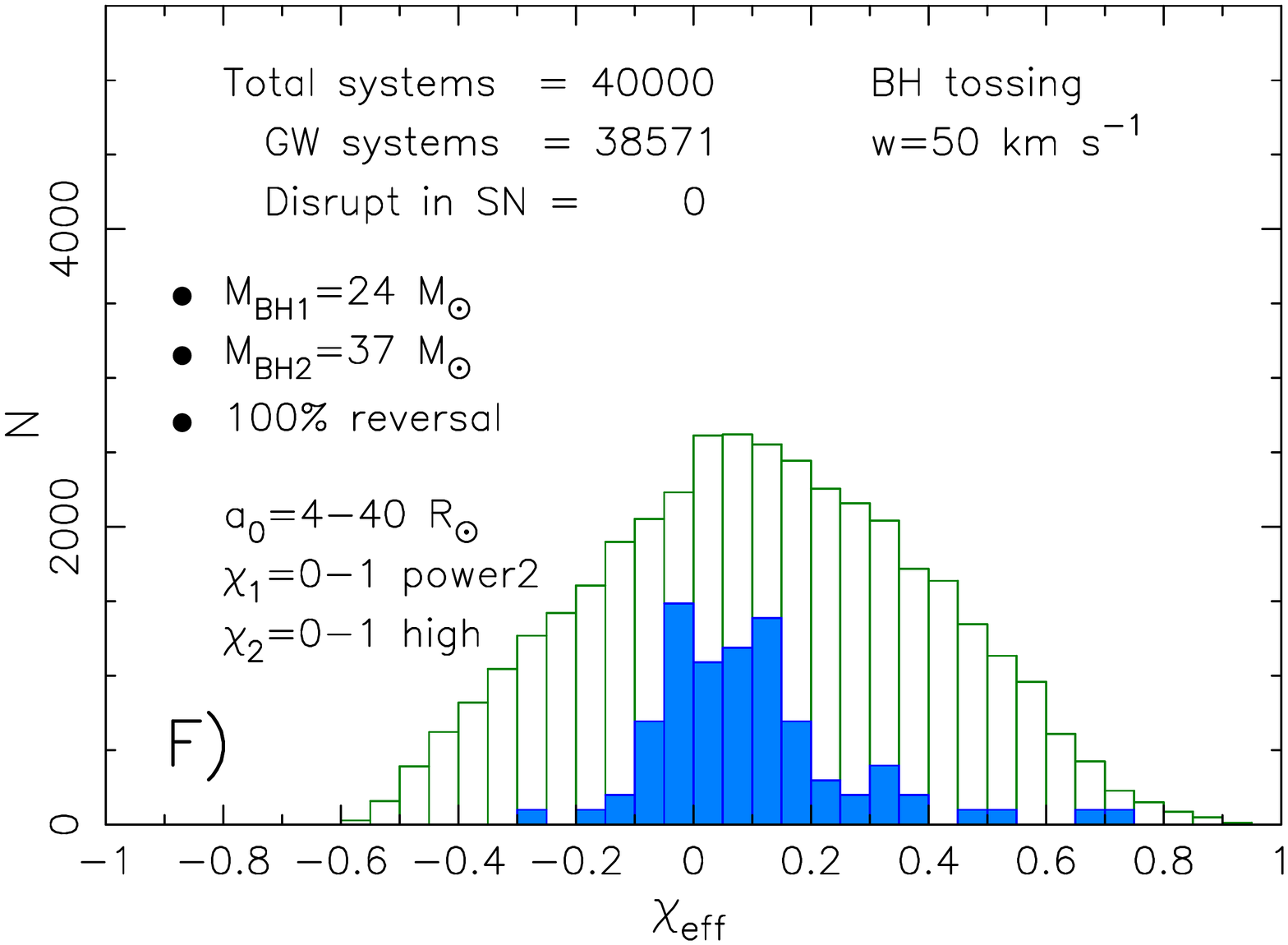}

\vspace*{-0.3cm}
\caption{Distributions of simulated $\chi_{\rm eff}$ values as a function of applying different pre-SN orbital separations, $a_0$ ($4\;R_\odot$ and $40\;R_\odot$, top panels) and masses of the two resulting BH components, $(M_{\rm BH,1}/M_\odot,\,M_{\rm BH,2}/M_\odot)=(12,\,8), (59,\,16), (37,\,24), (24,\,37)$ (central and bottom panels). 
The $\chi_{\rm eff}$ distribution is basically invariant to $a_0$ (except that fewer systems merge within a Hubble time if $a_0=40\;R_\odot$) whereas it is slightly affected by the BH component masses.
\label{fig:panel6-2}}
\end{figure*}

\begin{figure*}
\vspace*{-1.0cm}
\hspace*{-0.8cm}
\includegraphics[width=0.58\textwidth]{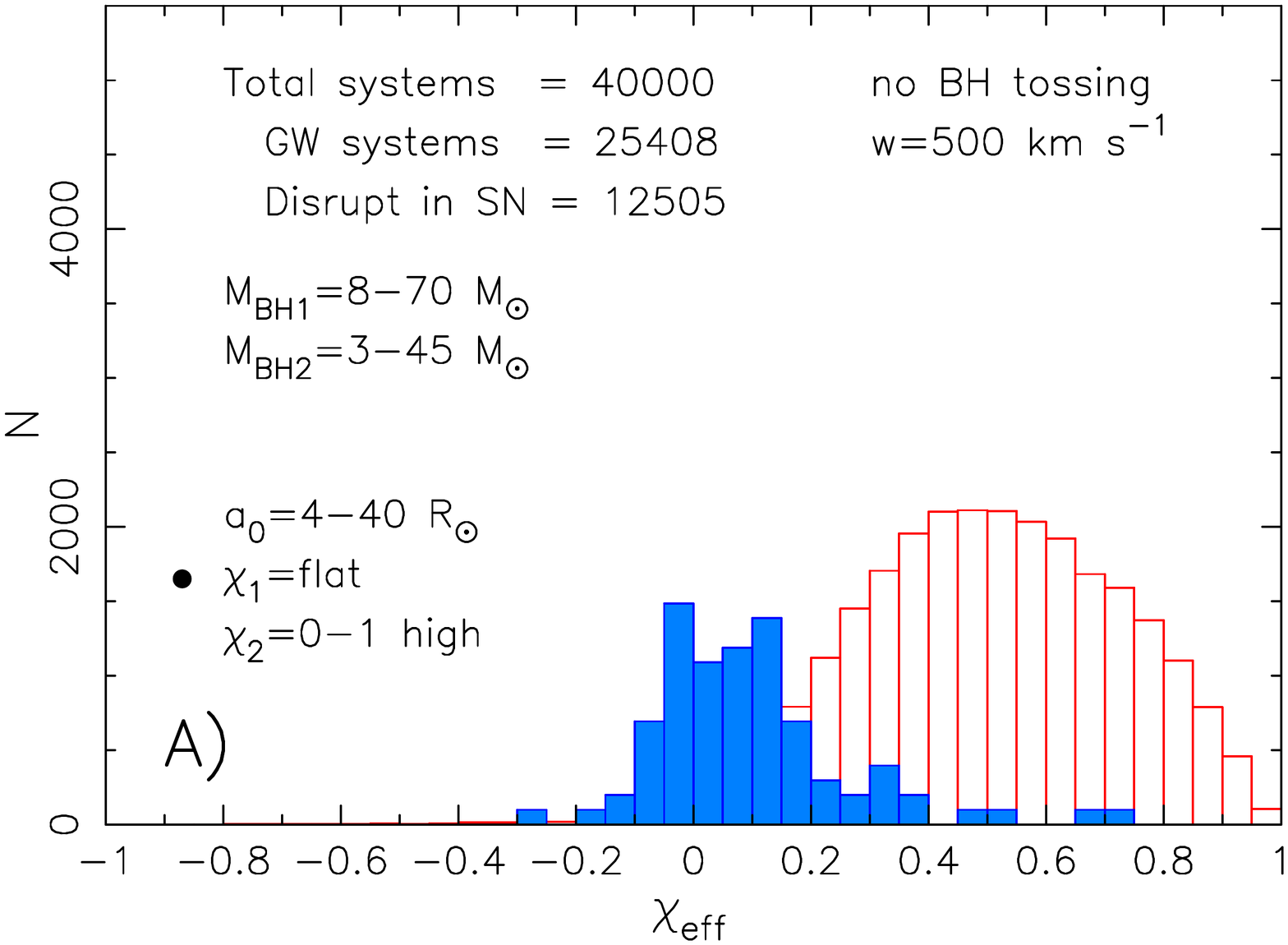}
\hspace*{-1.3cm}
\includegraphics[width=0.58\textwidth]{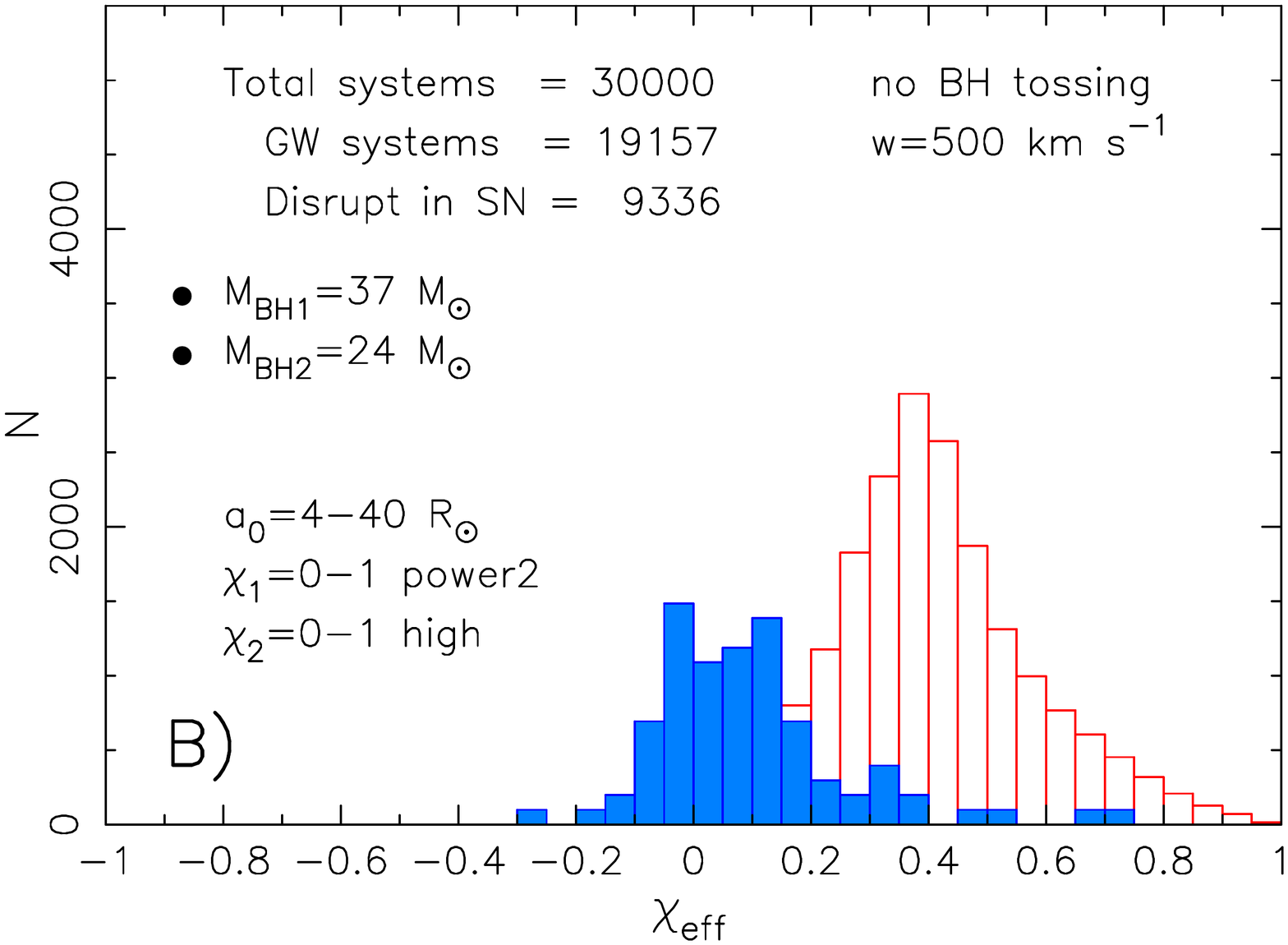}

\vspace*{-1.2cm}
\hspace*{-0.8cm}
\includegraphics[width=0.58\textwidth]{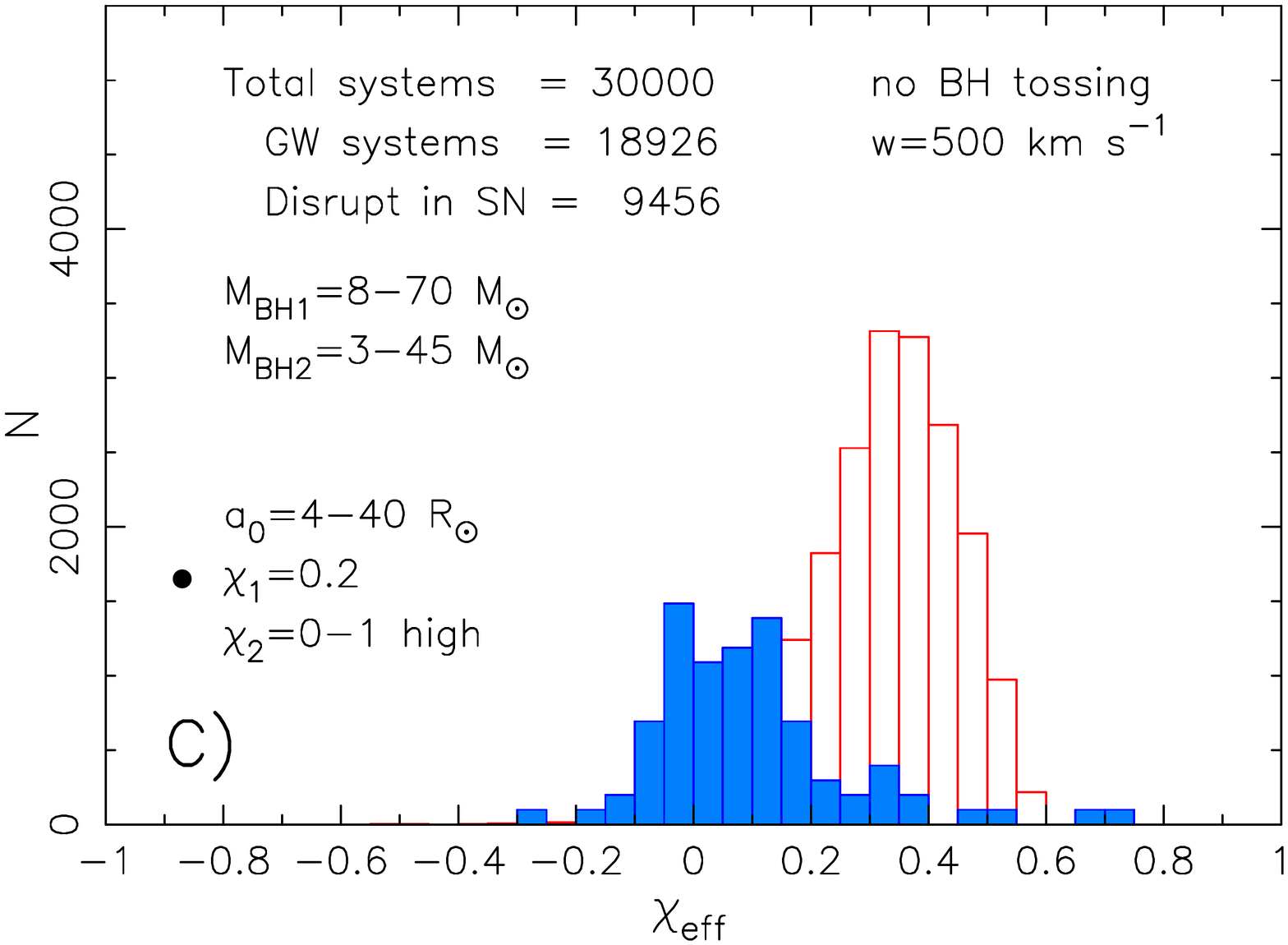}
\hspace*{-1.3cm}
\includegraphics[width=0.58\textwidth]{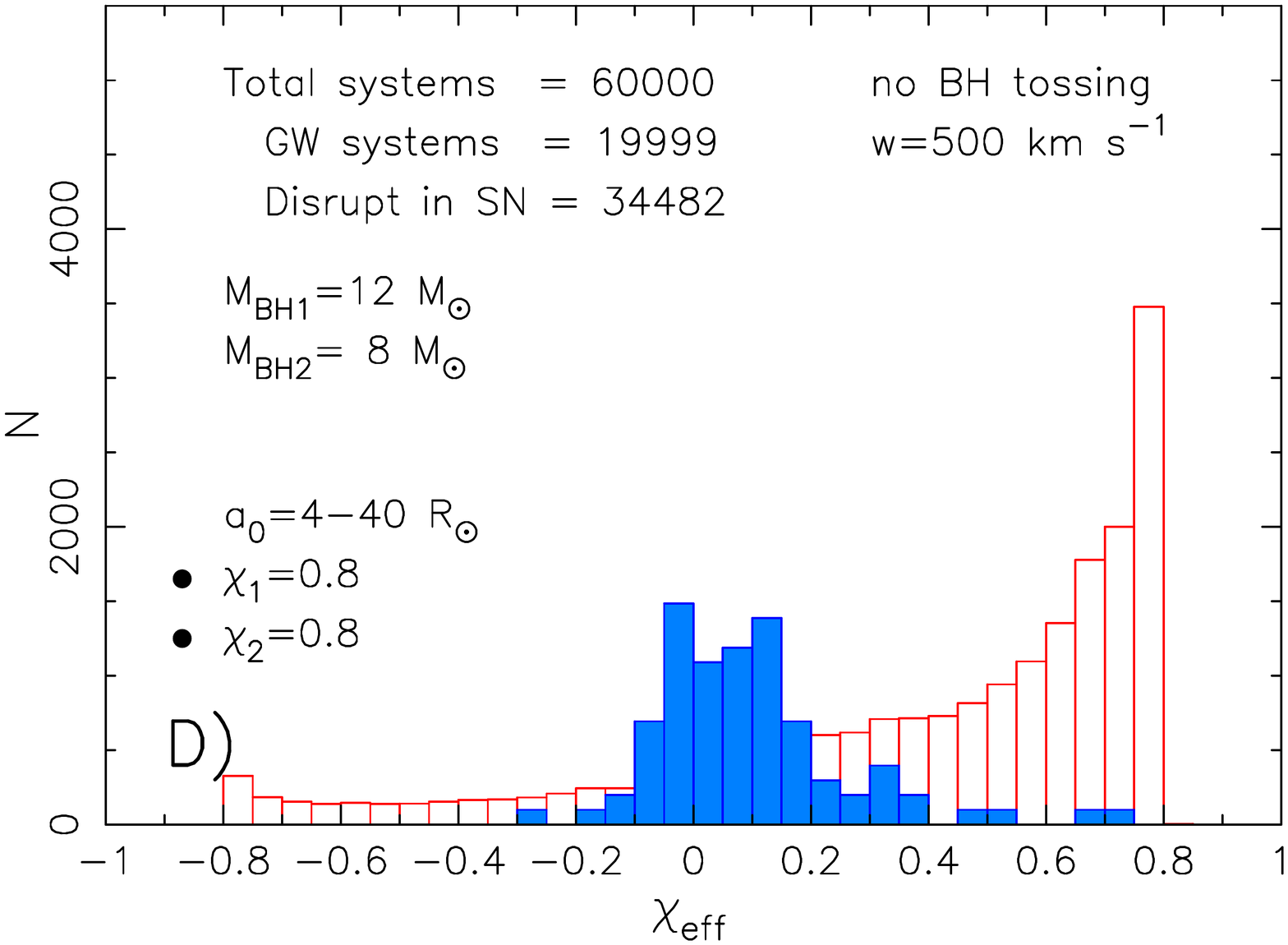}

\vspace*{-1.2cm}
\hspace*{-0.8cm}
\includegraphics[width=0.58\textwidth]{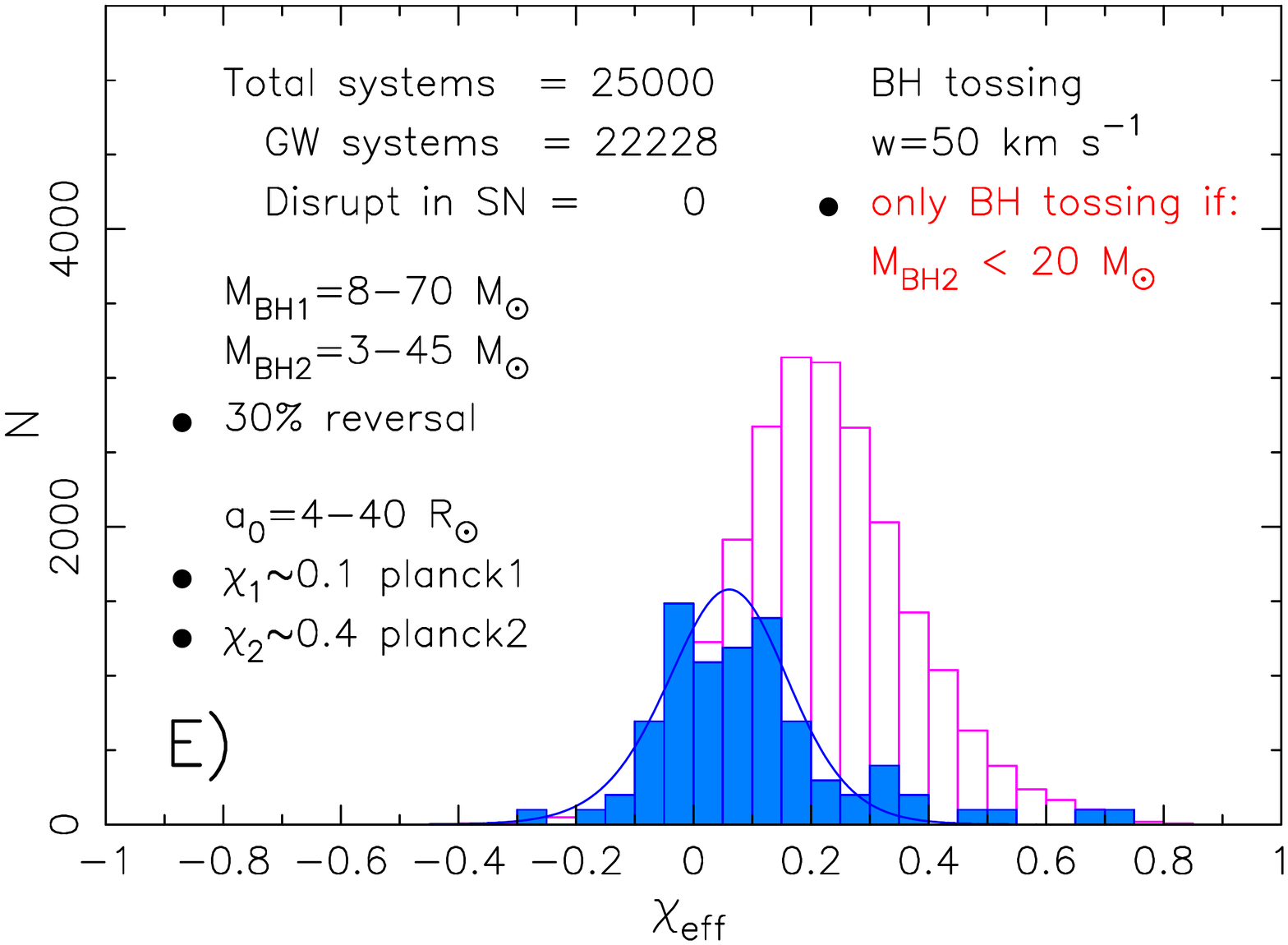}
\hspace*{-1.3cm}
\includegraphics[width=0.58\textwidth]{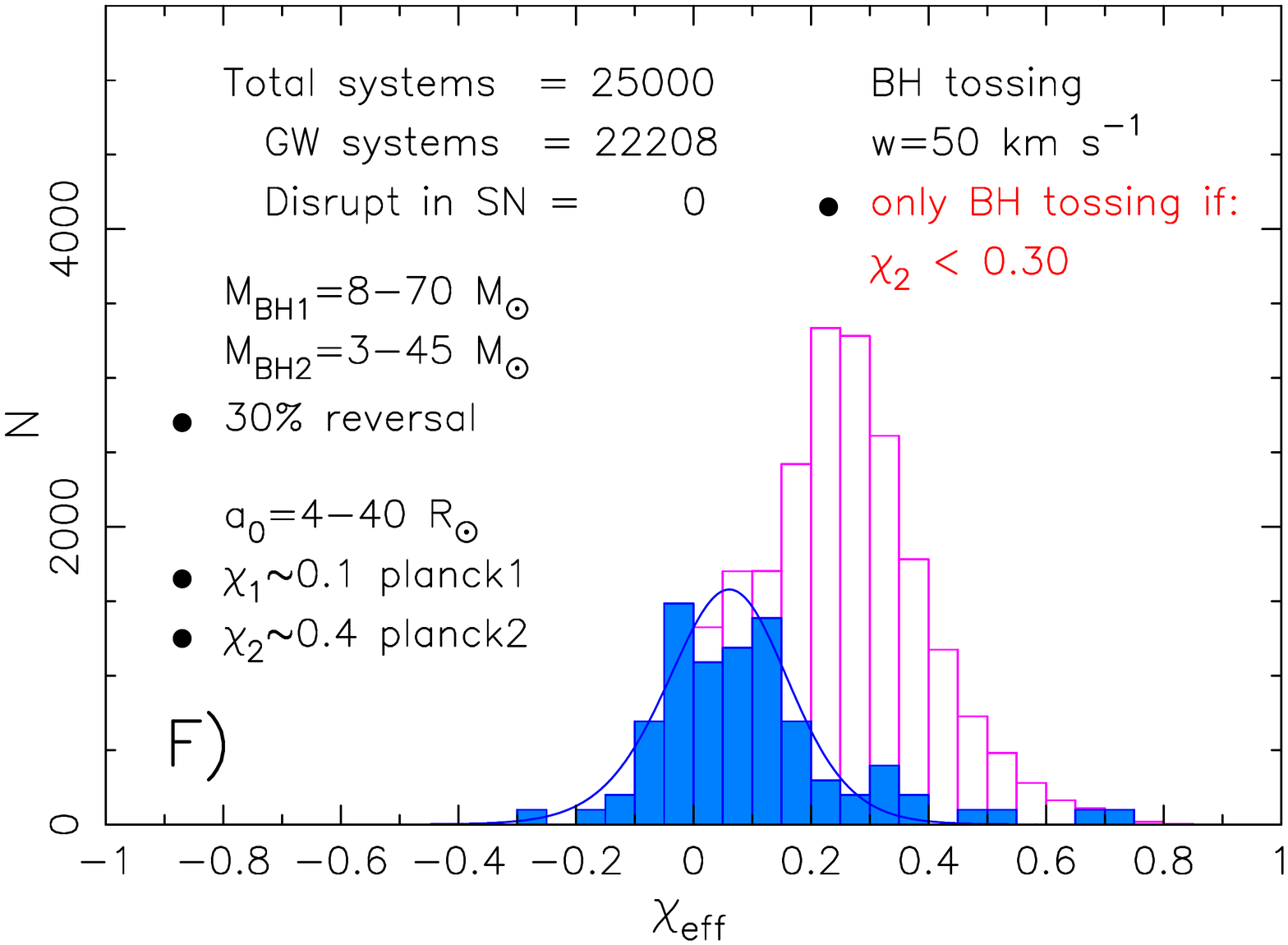}

\vspace*{-0.3cm}
\caption{Distributions of simulated $\chi_{\rm eff}$ values as a function of applying different BH component spin distributions ($\chi_1$ and $\chi_2$), in panels A--D excluding BH spin-axis tossing and assuming a kick of $w=500\;{\rm km\,s}^{-1}$. Even such large kicks cannot shift the distributions to sufficiently small values of $\chi_{\rm eff}$ in general; only low-mass BH+BH systems (panel~D) allow significant cases with $\chi_{\rm eff}<0$.
Panels E and F only include BH spin-axis tossing if $M_{\rm BH,2}<20\;M_\odot$ or if $\chi_2<0.30$, respectively (see text).
\label{fig:panel6-3}}
\end{figure*}

\begin{figure*}
\vspace*{-1.0cm}
\hspace*{-0.8cm}
\includegraphics[width=0.58\textwidth]{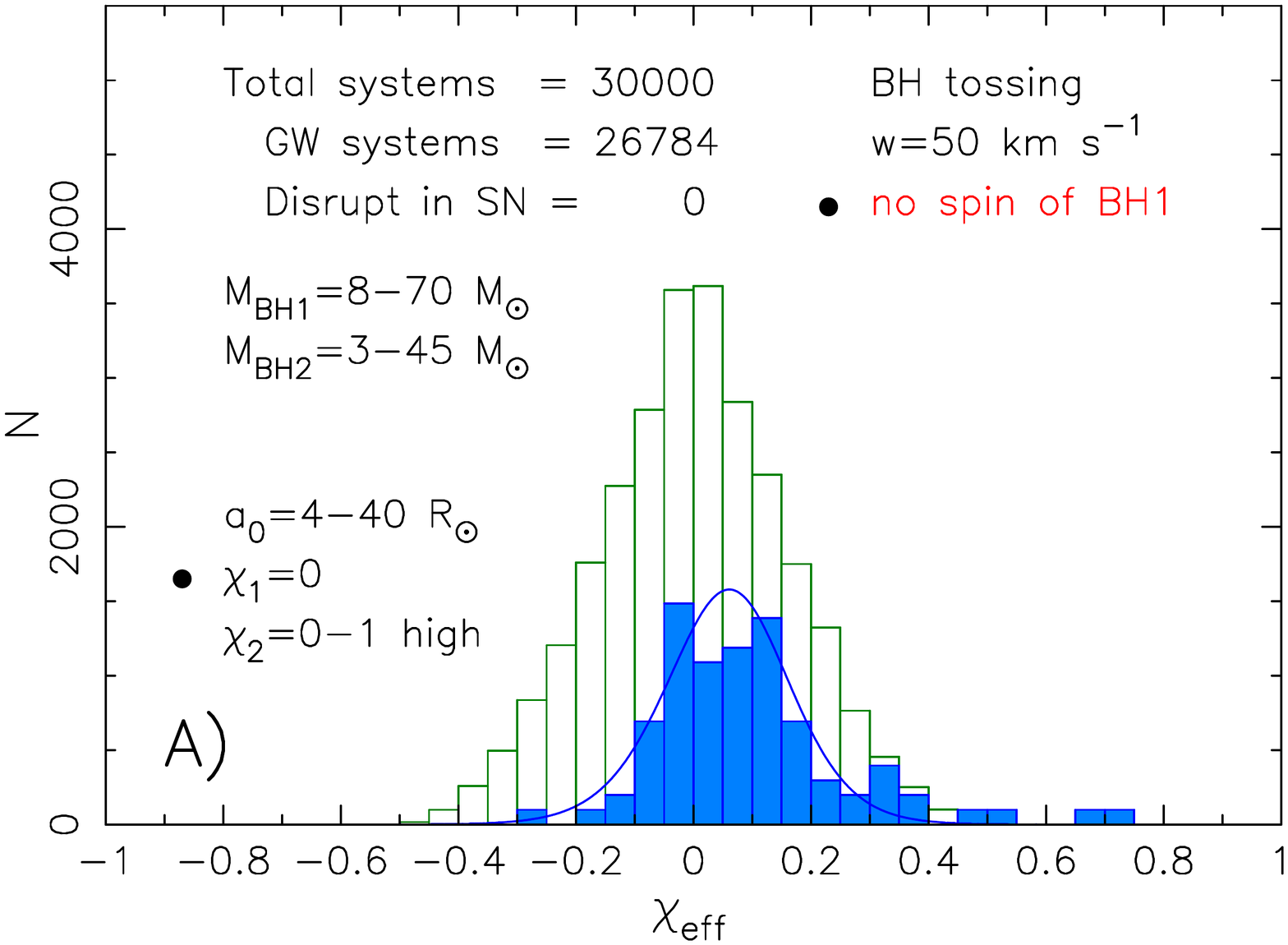}
\hspace*{-1.3cm}
\includegraphics[width=0.58\textwidth]{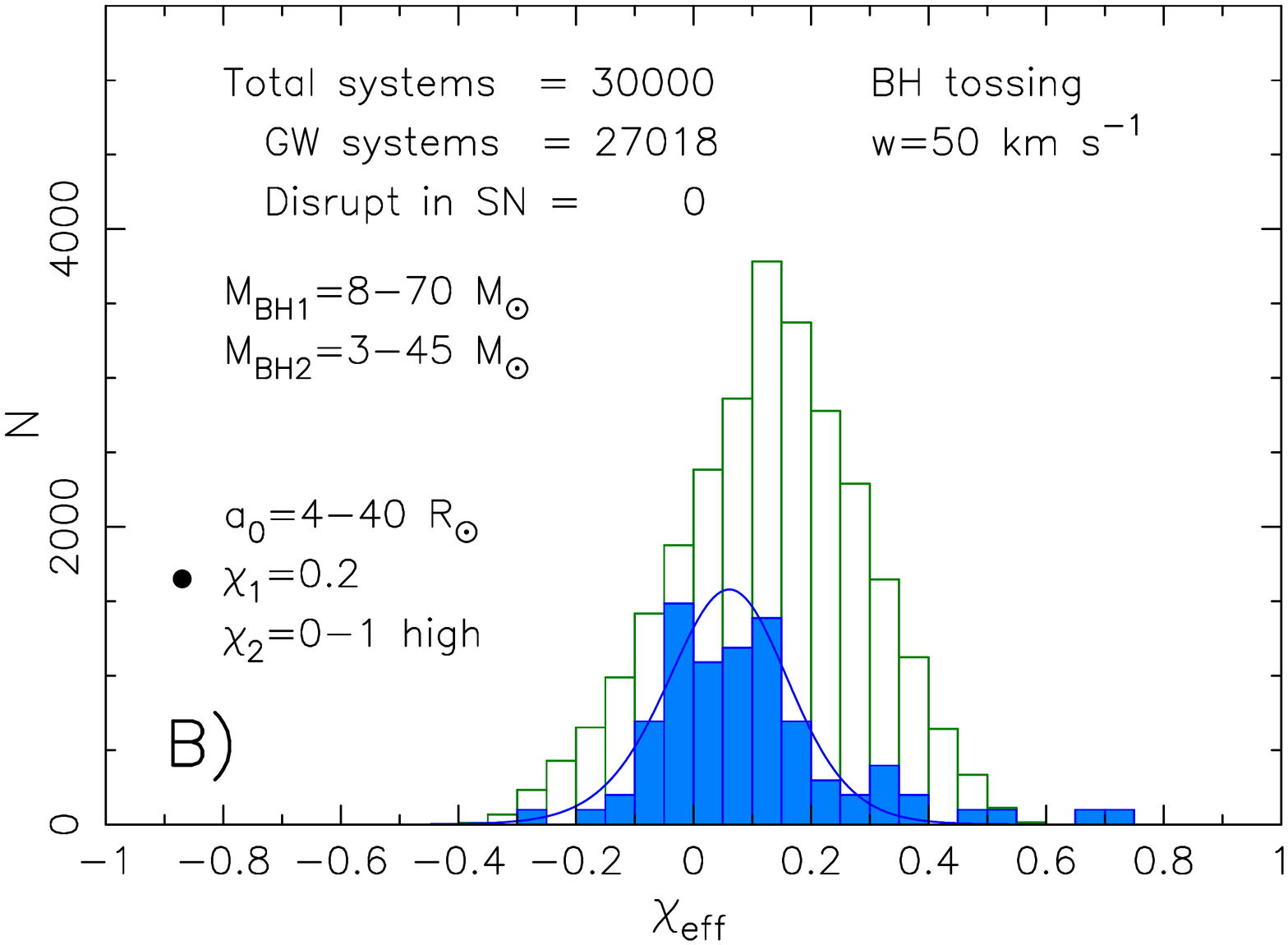}

\vspace*{-1.2cm}
\hspace*{-0.8cm}
\includegraphics[width=0.58\textwidth]{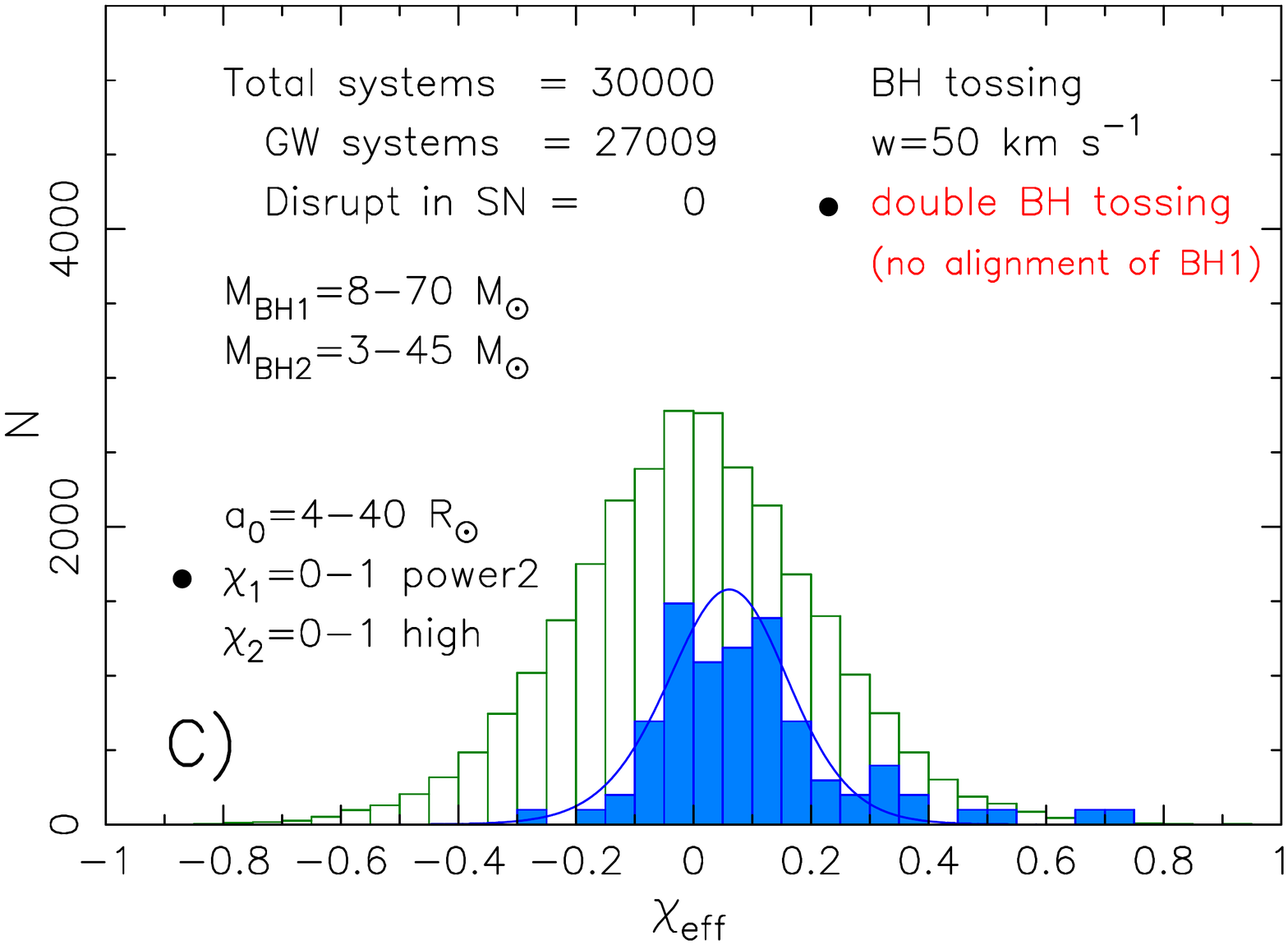}
\hspace*{-1.3cm}
\includegraphics[width=0.58\textwidth]{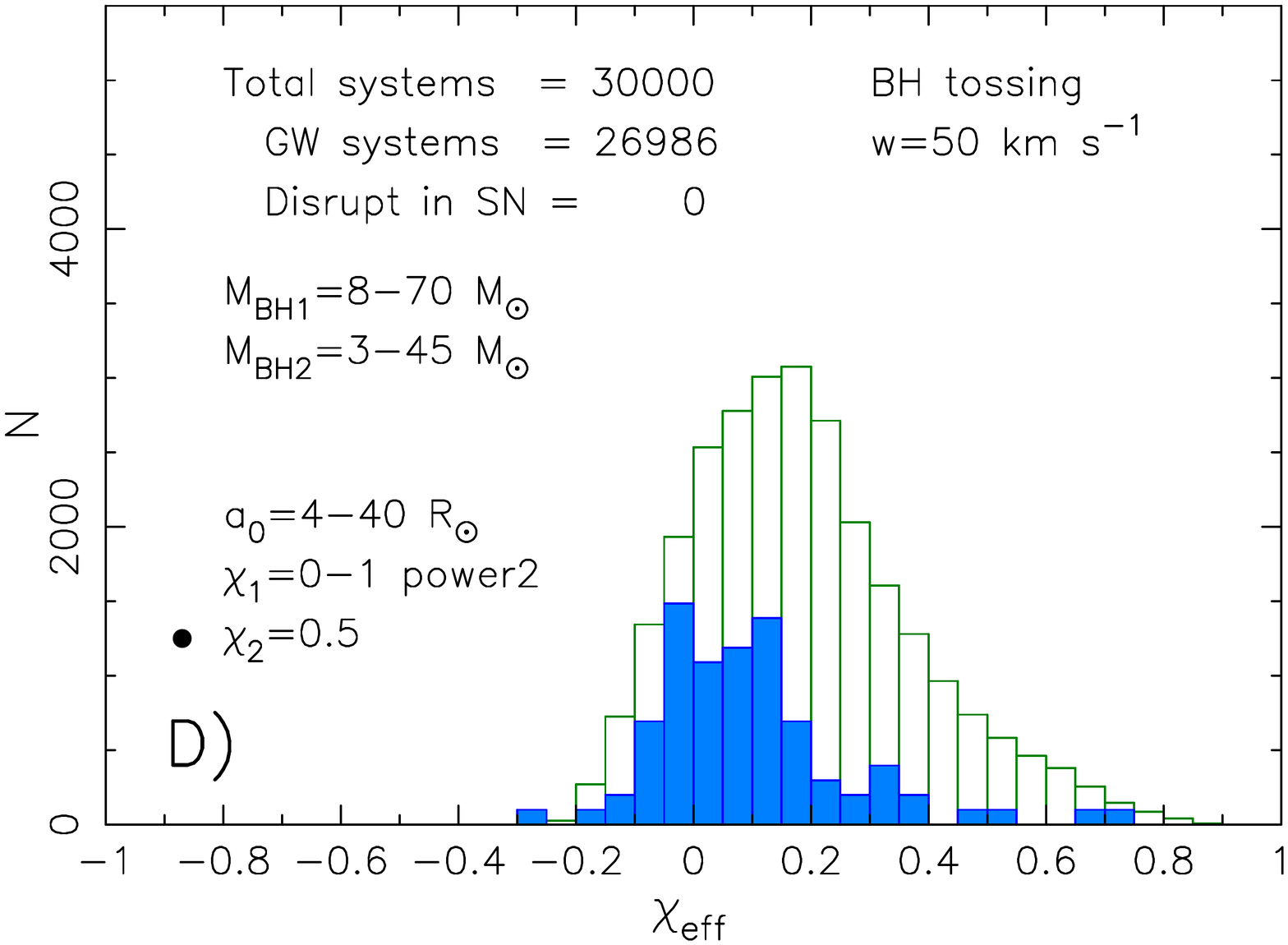}

\vspace*{-1.2cm}
\hspace*{-0.8cm}
\includegraphics[width=0.58\textwidth]{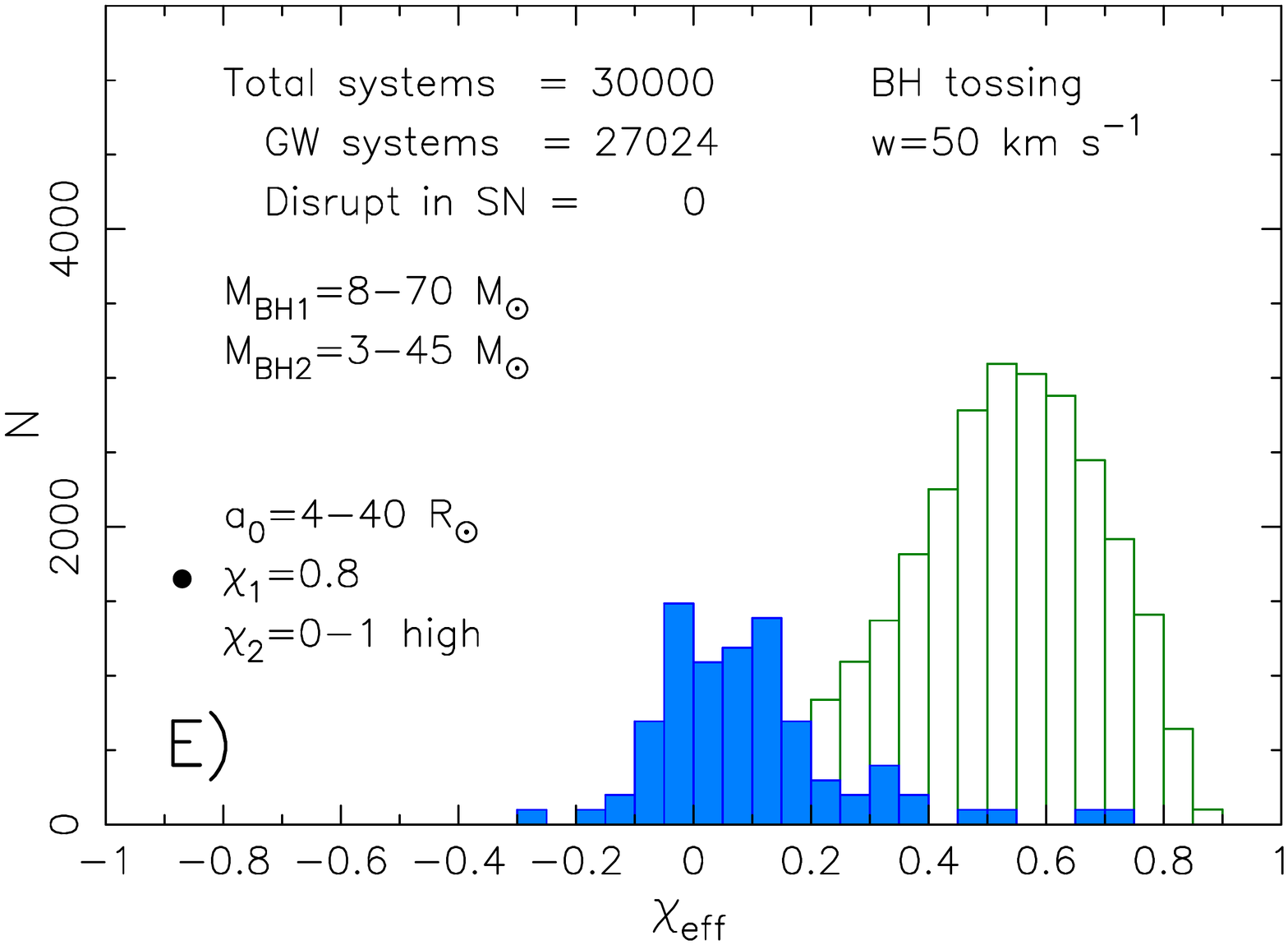}
\hspace*{-1.3cm}
\includegraphics[width=0.58\textwidth]{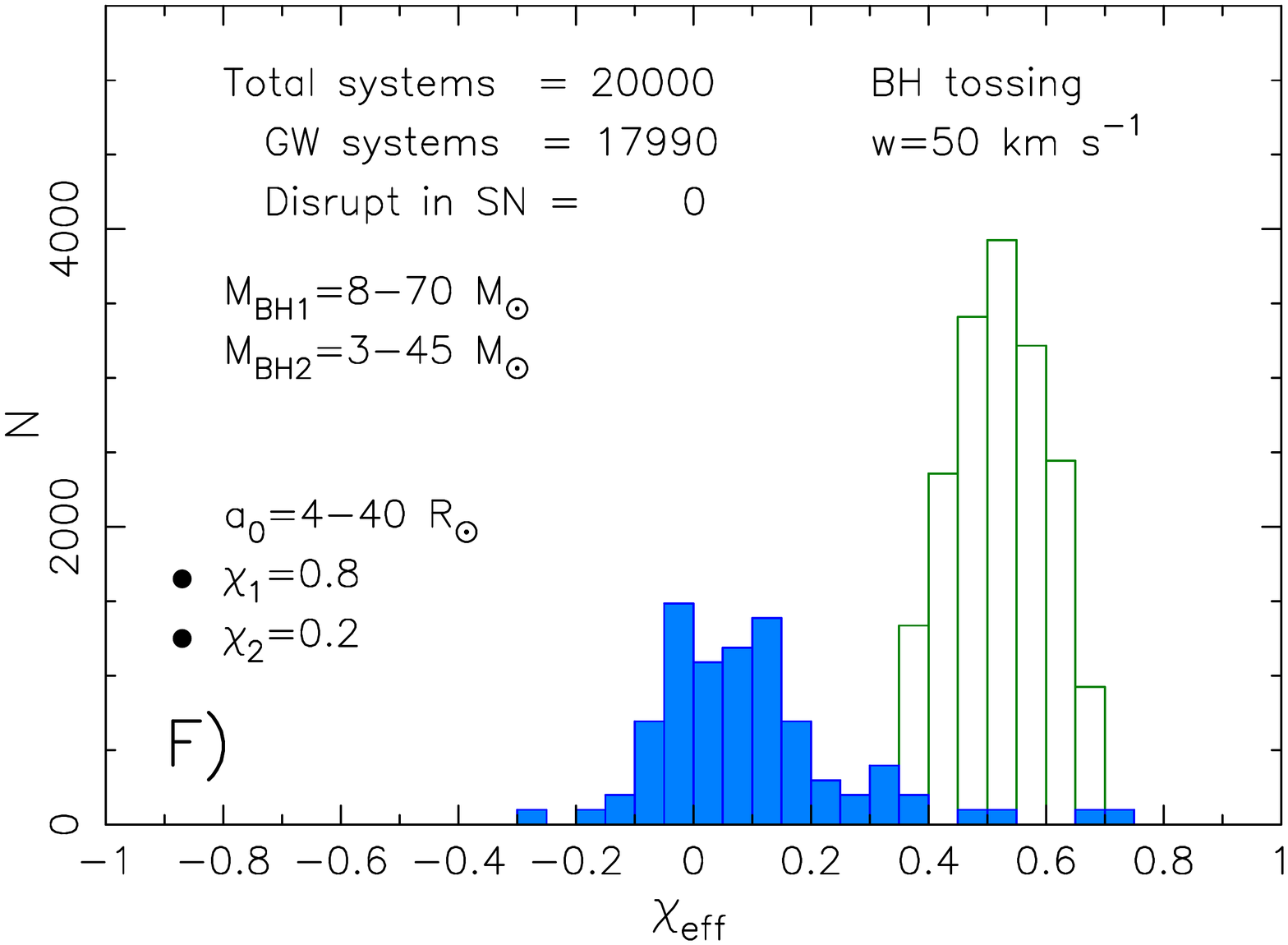}

\vspace*{-0.3cm}
\caption{Distributions of simulated $\chi_{\rm eff}$ values as a function of applying different BH component spin distributions ($\chi_1$ and $\chi_2$). 
In general, the fit to the empirical data is best for small values of $\chi_1\sim 0.1$ and medium to large values of $\chi_2$, in accordance with expectations from binary evolution theory (see text, and compare with histograms in other figures). Panels~A ($\chi_1=0$) and C (no alignment of first-born BH, i.e. in effect ``double'' BH tossing) have fully symmetric $\chi_{\rm eff}$ distributions as expected.
\label{fig:panel6-4}}
\end{figure*}

\begin{figure*}
\vspace*{-1.0cm}
\hspace*{-0.8cm}
\includegraphics[width=0.58\textwidth]{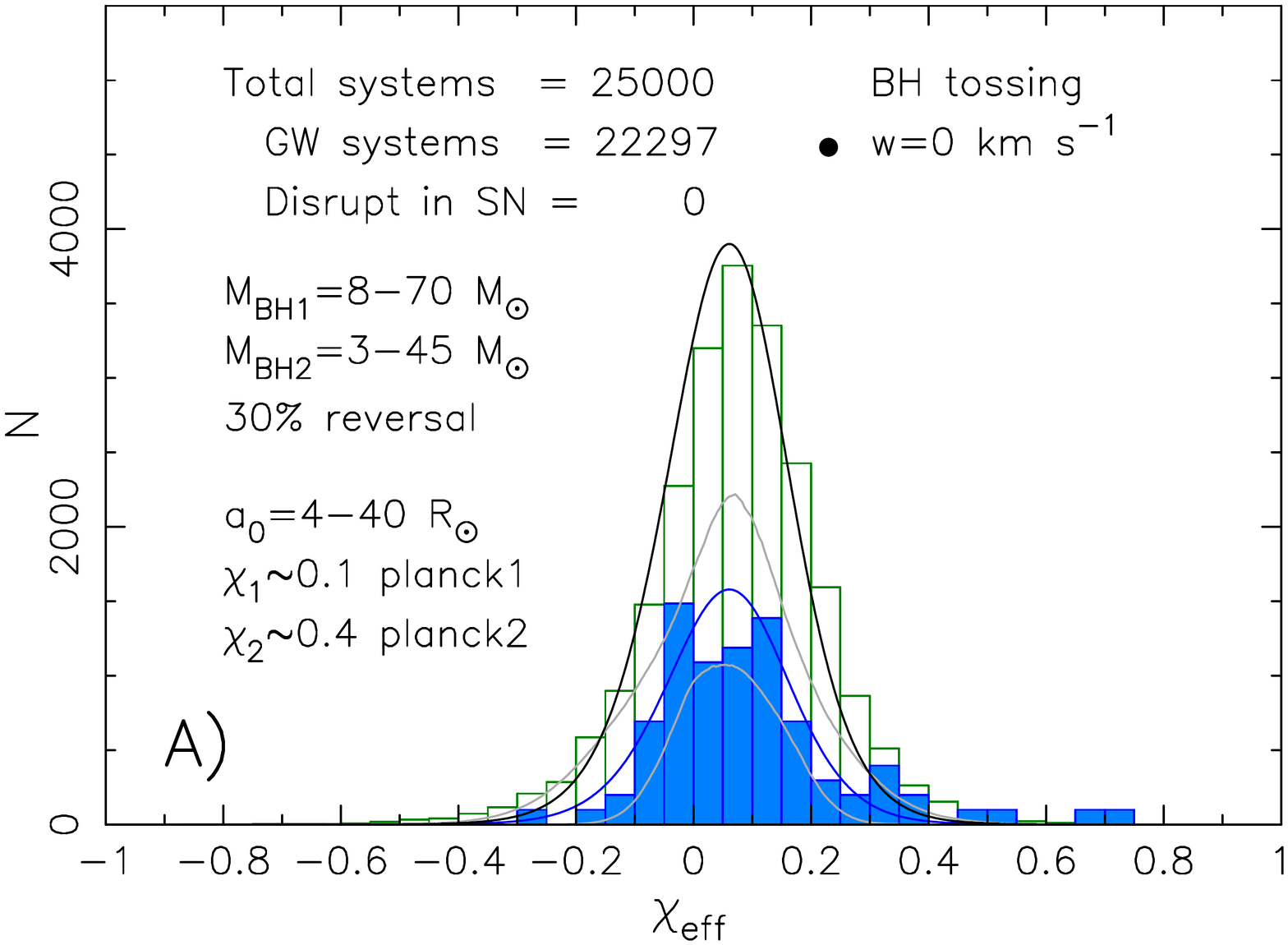}
\hspace*{-1.3cm}
\includegraphics[width=0.58\textwidth]{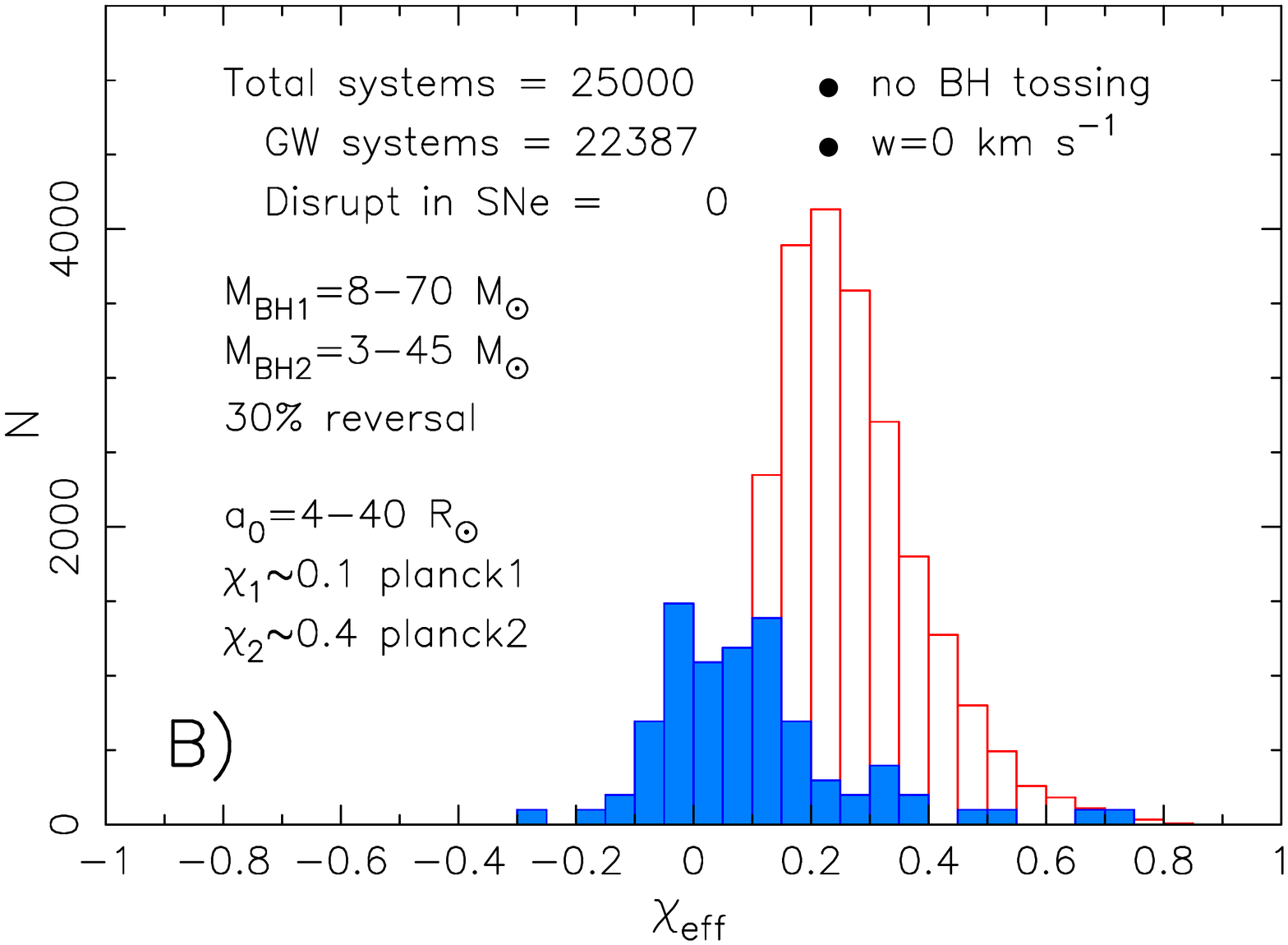}

\vspace*{-1.2cm}
\hspace*{-0.8cm}
\includegraphics[width=0.58\textwidth]{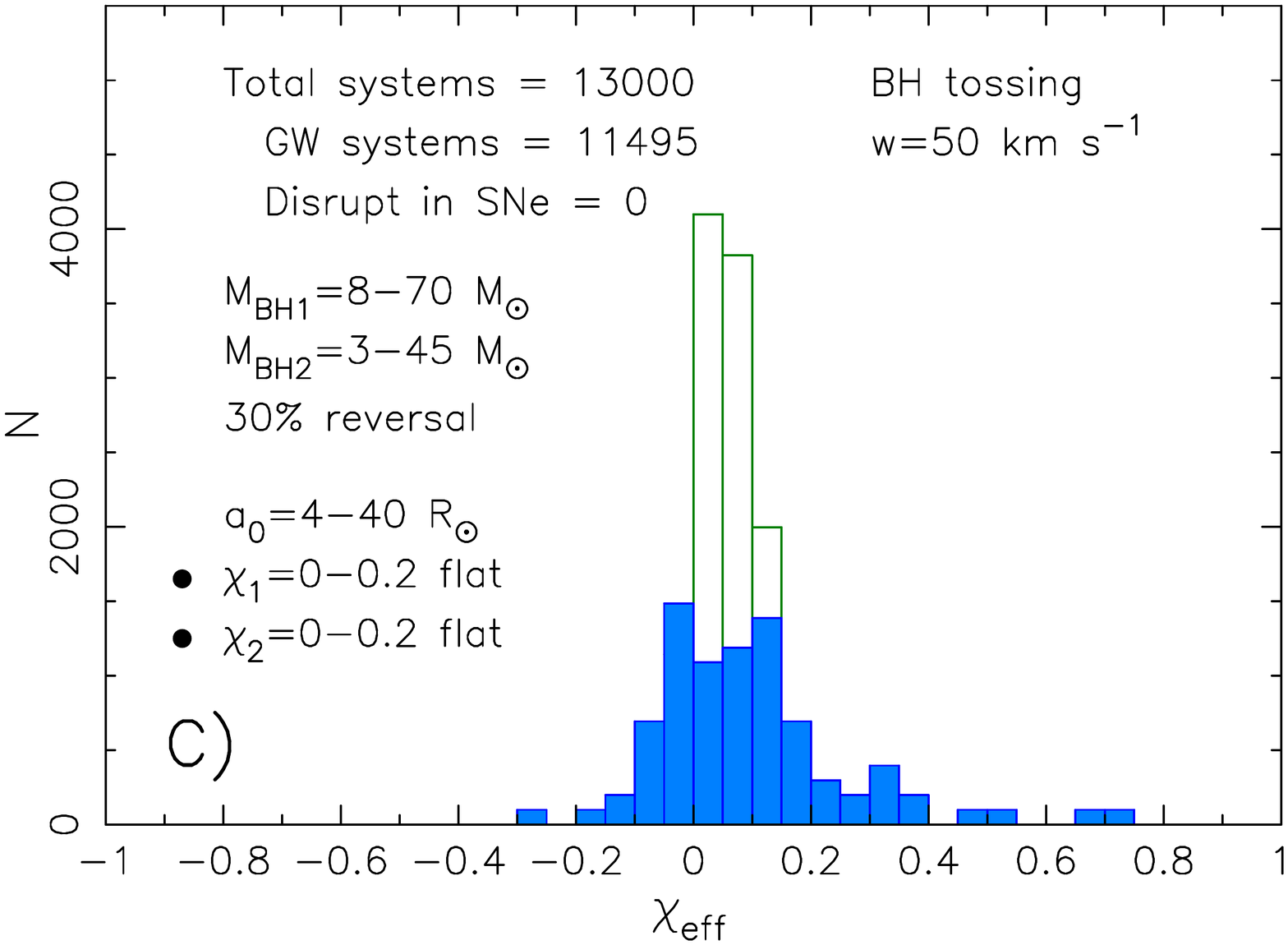}
\hspace*{-1.3cm}
\includegraphics[width=0.58\textwidth]{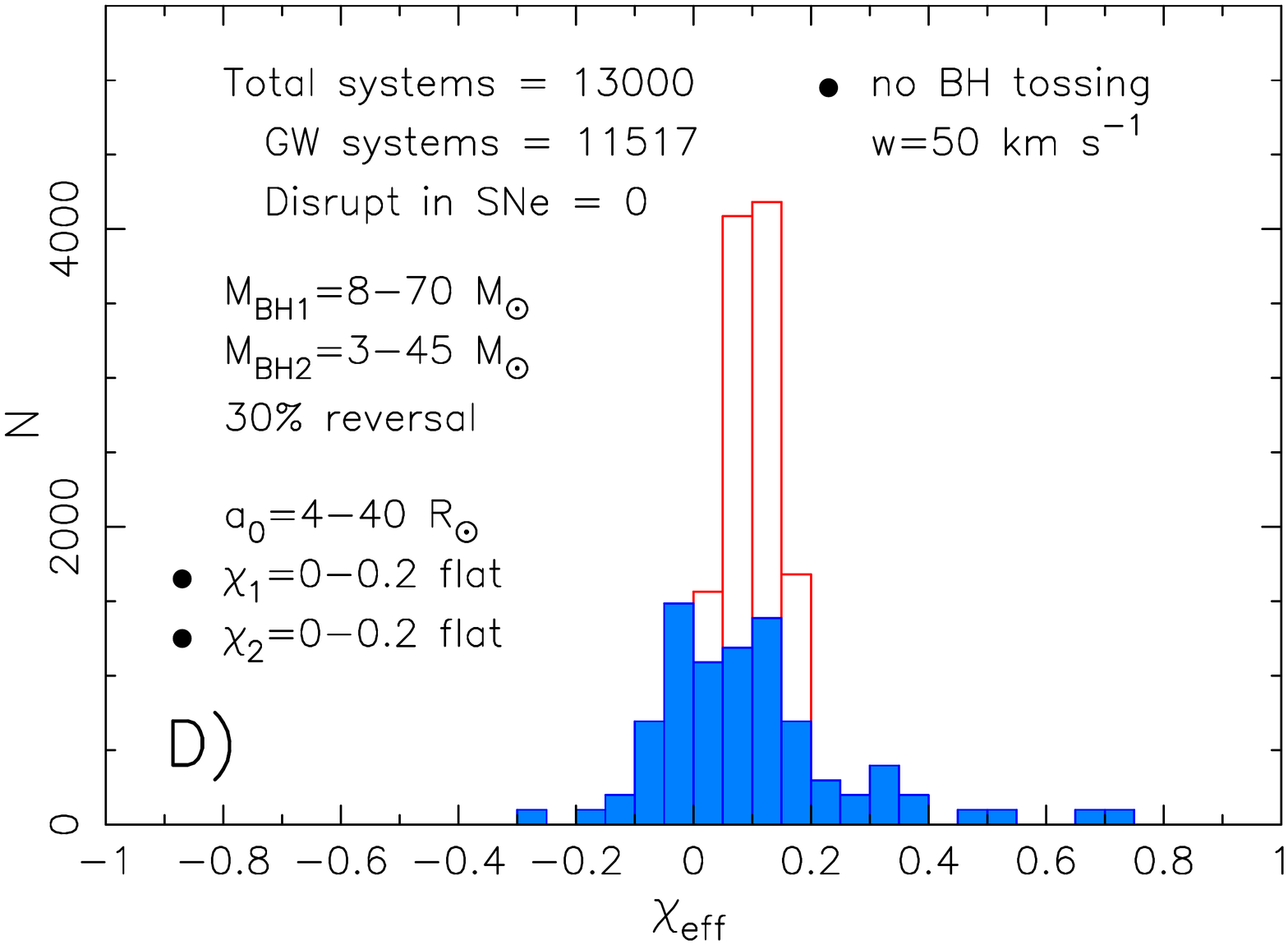}

\vspace*{-1.2cm}
\hspace*{-0.8cm}
\includegraphics[width=0.58\textwidth]{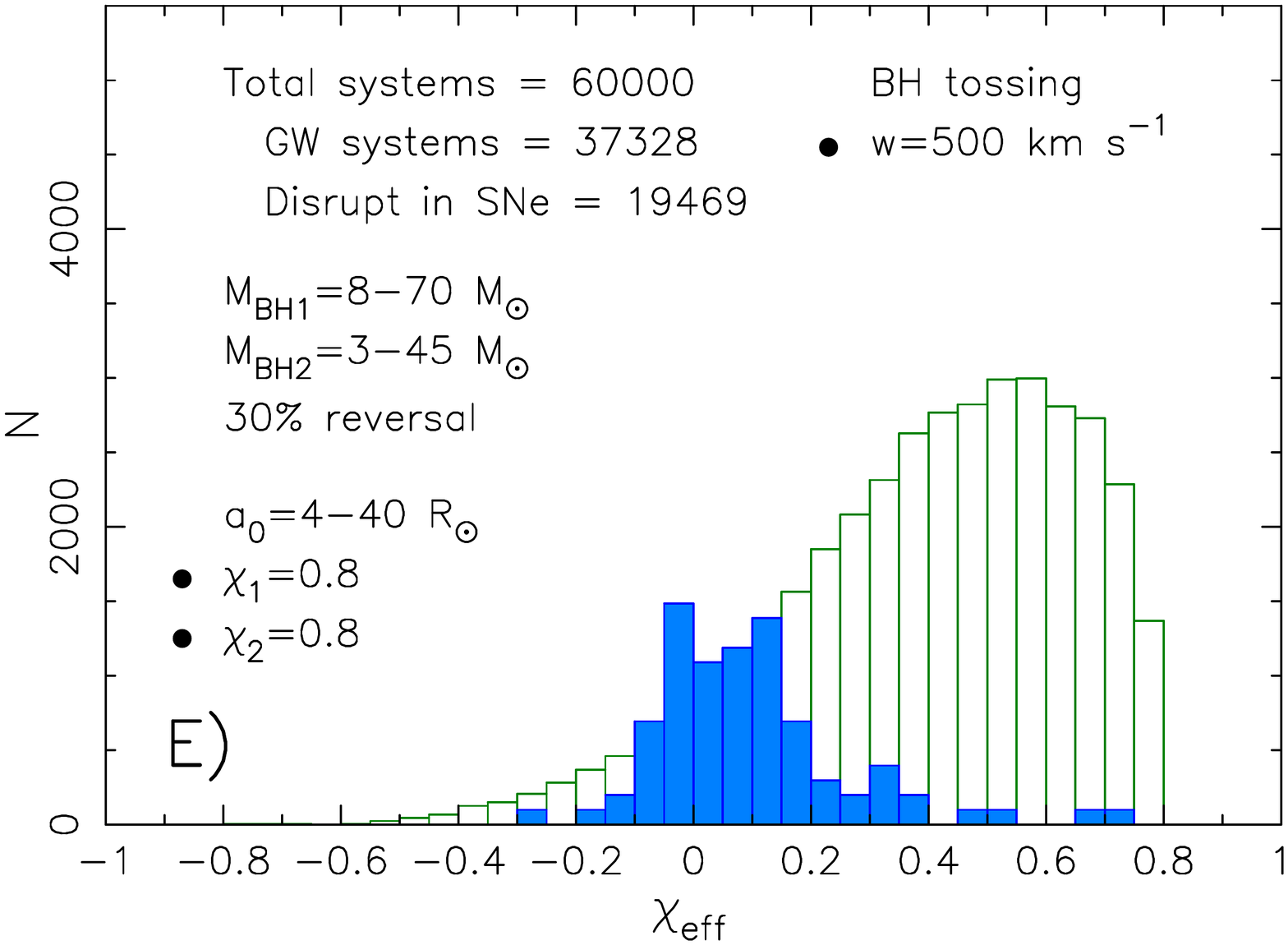}
\hspace*{-1.3cm}
\includegraphics[width=0.58\textwidth]{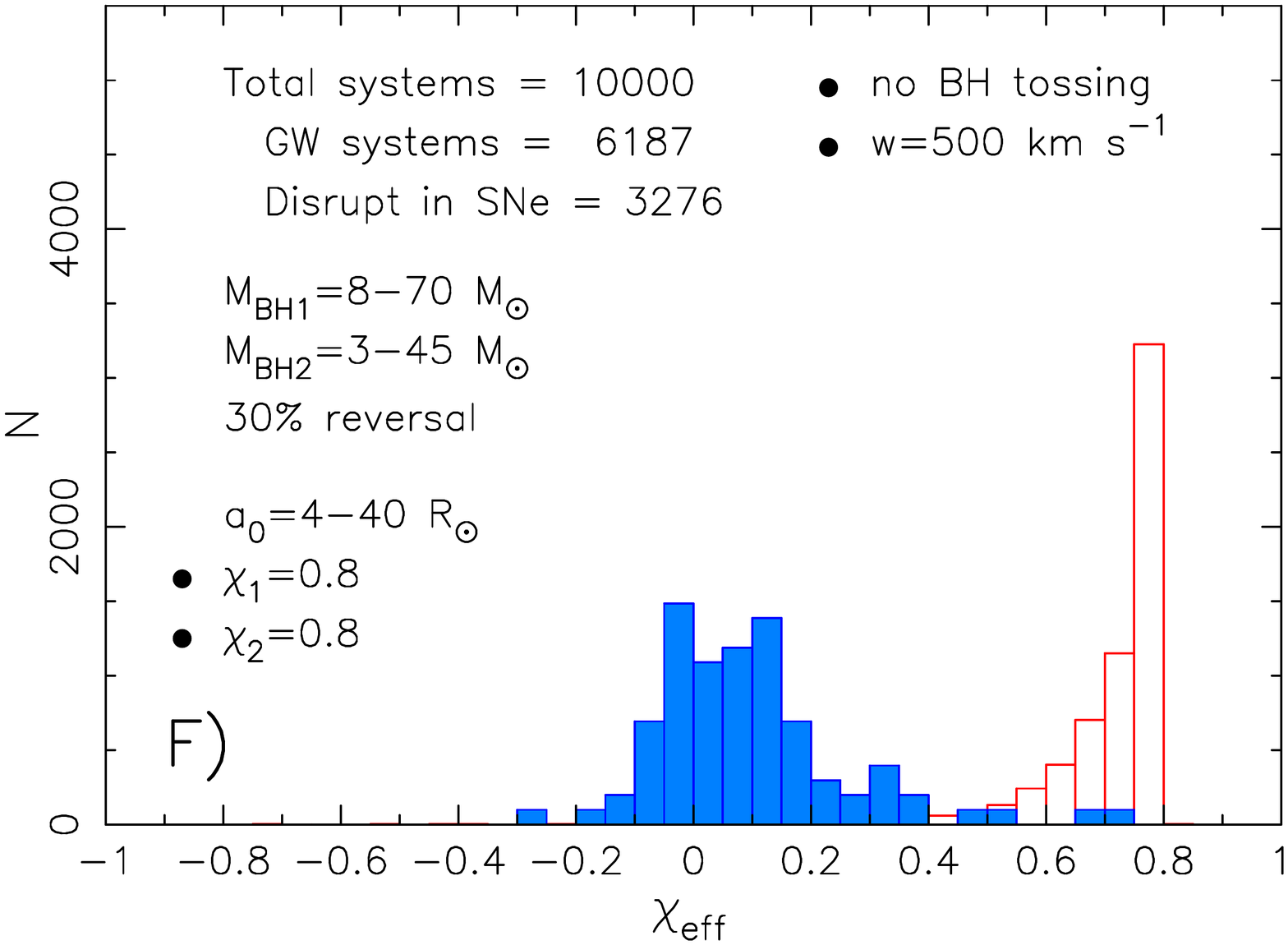}

\vspace*{-0.3cm}
\caption{Distributions of simulated $\chi_{\rm eff}$ values assuming no kicks ($w=0$, top row) or as a function of exploring different BH component spin values ($\chi_1$ and $\chi_2$, panels~C to F). Left column shows simulations including BH tossing; right column is excluding BH tossing (see text). 
In panel~A, the curves are: (blue) the mean estimate of the PDF of $\chi_{\rm eff}$ from GWTC-3 data \citep{aaa+22b}, (grey) the 5\% and 95\% quantiles, and (black) the blue curve normalized to number of simulated systems.
\label{fig:panel6-5}}
\end{figure*}

If the far majority of the observed BH+BH systems do have an origin from isolated binaries (as supported by simulations presented here), a strong mismatch between observations and simulations is obtained in cases where the first-born BH is assumed to be fairly rapidly spinning (see e.g. panels~E and F in Fig.~\ref{fig:panel6-4}). This result is in agreement with expectations from binary stellar evolution theory \citep{hws05,fm19} predicting a small spin for the first-born BH ($\chi_1\lesssim 0.1$). 
Indeed, it is demonstrated here (Fig.~\ref{fig:panel6} [panel~A] and Fig.~\ref{fig:panel6-5} [panel~A], both applying ``planck1'' and ``planck2'' spin PDFs for $\chi_1$ and $\chi_2$, see Table~\ref{table:chi} and Fig.~\ref{fig:chi_dist}) that individual spin magnitudes of order $\chi_1\sim 0.1$ and $\chi_2\sim 0.4$ reproduce the empirical data quite well (see Section~\ref{subsec:BH_component_spins} for further discussion). This is in agreement with a statistical analysis of the currently measured data \citep[see][and their fig.~17]{aaa+22b}. 
It is also seen from the empirical data that the average sum of the two spin magnitudes should not be too large ($|\chi_1|+|\chi_2|\lesssim 0.8$). 

Obviously, one can obtain desired final $\chi_{\rm eff}$ values in a narrow band by selecting $\chi_1$ and $\chi_2$ values in a narrow interval and applying small kicks (see e.g. panels~C and D in Fig.~\ref{fig:panel6-5}). 

\section{Discussions}\label{sec:discussions}

\subsection{BH component spins and alignment}\label{subsec:BH_component_spins}
The spin magnitude of a newborn BH reflects the spin of the core of its collapsing progenitor star, which itself depends on the angular momentum transport during its previous phases of stellar evolution, including stellar winds and accretion. Efficient angular momentum transport by viscosity and magnetic torques will couple the stellar core to its envelope, thereby slowing the spin of the core as the envelope expands when it becomes a giant star. Based on such studies, it has been argued \citep{hws05,fm19} that BHs from single stars, and first-born BHs in binaries, are expected to have very slow spins. This motivates the green (``planck1'') and blue (``power2'') PDFs of $\chi_1$ shown in Fig.~\ref{fig:chi_dist} (see Table~\ref{table:chi} for details). 

The rationale for allowing the possibility $\chi_1\rightarrow 1$ is twofold: first, the estimated spins of BHs observed in HMXBs are apparently all quite large \citep{mbo+21} with values between $0.84-0.99$; and, second, some of the most massive BH components (and thus in general expected to have formed first) in recorded BH+BH mergers seem to possess rather large values of $\chi_1$ \citep[][with the usual caveat that these measurements have large uncertainties in general]{aaa+22b}. 
This evidence suggests that either: i) isolated binary evolution may produce larger spins of the first-born BHs than predicted from current theory, ii) the BHs in these HMXBs and BH+BH merger sources could possibly have obtained their rapid spins from chemical homogeneous evolution \citep{mlp+16,md16b}, or iii) the general populations of \mbox{(near-)} Galactic HMXBs with rapid BH spins and BH+BH mergers are distinct \citep{gfk+22}. The rapidly spinning BHs in the former population may give support to the third possibility and could therefore conveniently alleviate the problem of aligning rapidly spinning BHs (see below), following a core collapse with spin axis tossing. 

Rapid BH spins obtained from Eddington-limited accretion is not possible within the short lifetime of the their massive companion stars. From similar simple arguments as in e.g. \citet{mf20}, one can show that the ratio of orbital angular momentum of an accreted test particle (with mass $m$, carrying the specific orbital angular momentum from the innermost stable circular orbit, ISCO) and the spin angular momentum of a slow-spinning accreting BH (with mass $M$) is given by:
\begin{equation}
\frac{J}{J_{\rm BH}}\simeq \frac{1}{\chi}\;\frac{m}{M}\sqrt{\frac{R_{\rm ISCO}\,c^2}{GM}} \sim \frac{\sqrt{12}}{\chi}\;\frac{m}{M}
\end{equation}
Hence, for $\chi \simeq 0.1$, we see that $J/J_{\rm BH}\sim 35\;m/M$, and thus accreting an amount of mass equivalent to a few percent of the BH can make a significant contribution to $J_{\rm BH}$, and therefore likely enable alignment (at least for slow-spinning BHs). 

Moreover, and more important, tides \citep{hut81} and torques from BH--disk interactions \citep[Bardeen-Petterson effect,][]{bp75,np98,klop05} have been shown to align the BH spin with the orbital angular momentum vector \citep[but see also][]{sa21}.
Finally, the subsequent common-envelope stage (central to most models of isolated binary star evolution) is poorly understood, and may also affect spin-orbit couplings of angular momenta from the bulk mass ejection on a short timescale. 

Panels E and F of Fig.~\ref{fig:panel6-3} were simulated with {\em partial} tossing, i.e. BH spin-axis tossing was implemented only for systems where $M_{\rm BH,2}<20\;M_\odot$ or $\chi_2<0.30$ (see Section~\ref{subsubsec:mechanism} for discussions). 
The resulting $\chi_{\rm eff}$ distributions do not match well with observations. This is not surprising since, in effect, these input conditions imply that a fairly large fraction of the simulations did not include tossing and therefore fails to reproduce the empirical data (cf. Fig.~\ref{fig:panel6}, panel~B).

Figure~\ref{fig:panel6-4} shows the effect of applying various BH spin magnitudes, in all cases including BH spin-axis tossing. 
Simulations with a constant large value of $\chi_1=0.8$ [panels~E,\,F] resulted in $\chi_{\rm eff}$ distributions that completely fail to explain the empirical data, thereby favouring smaller values of $\chi_1$ in general, if these systems formed from isolated binary star evolution. 
Panels~A and C serve as sanity checks: panel~A assumes $\chi_1=0$ (i.e. the first-born BH is non-spinning at the time of the second core collapse), and panel~C shows the effect of ``double'' BH tossing (i.e. equivalent to the situation in which no alignment occurs for the spin axis of the first-born BH during mass accretion from the companion star prior to the second SN). In both of these cases, we would expect the distribution of $\chi_{\rm eff}$ to be completely symmetric with respect to $\chi_{\rm eff}=0$. This is indeed in agreement with the presented simulated data, but in disagreement with observations, which is therefore another piece of evidence in favor of spin alignment prior to the second SN.

The spin of the second-born BH in an isolated binary system is also determined by angular momentum transport, stellar evolution and mass loss of its progenitor star. Particularly important for the second BH are tidal interactions in the post-HMXB (post-common envelope or post-RLO) phase where a naked helium star is orbiting the first-born BH in a tight orbit. Using analytical arguments and semi-analytical calculations, one can derive approximate estimates of the spin-up of the secondary naked core prior to its collapse and thereby constrain the spin of the second-born BH \citep{vy07,kzkw16,hp17,zkk18}. To mimick efficient spin-up via tides, PDFs ``high'' and ``planck2'' (see Table~\ref{table:chi}) were applied for $\chi_2$ in the work presented here.

\subsection{BH+BH outliers}\label{subsec:outliers}
Despite anticipated effective spin-up of the progenitor of the second-born BH due to tides \citep{kzkw16}, the two BH+BH mergers observed with large values of $\chi_{\rm eff}\sim 0.70^{+0.15}_{-0.25}$ require, in addition, a large spin of the first-born BH, thereby challenging theories on efficient angular momentum transport if they had formed through the isolated binary evolution channel \citep{qww+22}. One of these systems (GW190403) contains a very massive BH component ($88^{+38}_{-33}\;M_\odot$) that may alternatively suggest a product of one or more past mergers in a dense stellar environment \citep{rkg+20}, explaining both a large spin and a large mass.

\subsection{BH spin-axis tossing mechanism}\label{subsubsec:mechanism}
The new empirical data and the simulations presented here provide strong evidence for tossing of the spin axis during stellar collapse to a BH (as is evidently also the case when NSs form, see Section~\ref{sec:intro}). 
However, to produce a torque sufficiently strong to toss the spin axis during BH formation is not trivial. 

For NSs and low-mass BHs, the tossing mechanism (and remnant spin) may be caused by fallback of material on a timescale of minutes or even hours after the SN, as recently investigated by \citet{jwk21}. The tossing mechanism here is therefore a process distinct from that of the momentum kick imparted onto the remnant (on a timescale of a few seconds) due to mainly anisotropic mass ejection \citep{jan17} arising from non-radial hydrodynamic instabilities in the collapsing stellar core.
It is possible that BH kicks and spins are correlated in fallback scenarios \citep[e.g.][]{cmh20,jwk21}.

Another recent investigation by \citet{aq22} considers 3D convection in the hydrogen envelope of a massive progenitor star that is expected to collapse and directly produce a BH. They find that a very large amount of turbulent angular momentum is connected to convective mass motions. 
This means that when the BH is able to accrete only a fraction of the hydrogen envelope while the outer part of the envelope gets stripped according to the mass-decrement effect pointed out by \citet{nad80} and \citet{lw13}, the final BH spin can be rather large, even for a non-rotating pre-SN star, and the orientation of its spin axis will become random (i.e. random tossing of the BH spin axis as explored here). Whether or not this mechanism, which is a good candidate for a tossing scenario of massive BHs, also works efficiently in stripped envelope SNe remains to be proven. It is thus clear from the above two scenarios, that there could be a mass (and possibly spin) dependence on the BH tossing mechanism --- a feature that may be revealed from future data of BH+BH mergers.

\section{Conclusions}\label{sec:conclusions}
In this work, strong evidence has been presented for the first time for BH spin-axis tossing to operate during the formation of BHs in SNe. This finding is based on comparison of simulated distributions of $\chi_{\rm eff}$ of BH+BH systems produced in isolated binaries with empirical data of BH+BH mergers from GWTC-3 of the LIGO-Virgo-KAGRA network (up to and including O3a and O3b).
It is noteworthy that the theoretical simulations of the $\chi_{\rm eff}$ distribution are only weakly dependent on a large range of input parameters, which further strengthens the evidence for BH spin-axis tossing.  
Future comparison to empirical data should be followed up by further detailed and advanced statistical analysis, including selection effects, to rigidly test and confirm (or refute) the hypothesis presented here. The conclusions of this paper can be summarized as follows:

\begin{itemize}
    \item Empirical data (GWTC-3) strongly disfavor a main origin of BH+BH mergers from isolated binary star evolution {\em without} BH spin-axis tossing. Including (isotropic) BH tossing, however, provides a simple solution and an excellent fit to the bulk data for a large range of input distributions of relevant physical parameters (e.g. Fig.~\ref{fig:panel6}, panel~A). 
    
    \item An implication of this hypothesis is that it jeopardizes previous ideas on the use of BH spin data as a diagnostic for double BH formation models.
    
    \item The simulations presented here are broadly in agreement with empirically derived \citep{aaa+22b} spin magnitudes with average values of $\langle \chi_1 \rangle \sim 0.1$ and $\langle \chi_2 \rangle \sim 0.4$ for the first- and second-born BHs, respectively (cf. PDFs ``planck1'' and ``planck2'' in Fig.~\ref{fig:chi_dist}). However, we find that there is some indication that PDFs with a small tail extending up to $\chi_1 \rightarrow 1$, and increasing probability for larger $\chi_2$ values as expected from binary star interactions (PDFs ``power2'' ans ``high'', respectively), may fit the data even better.
    
    \item More statistical investigations (including selection effects) are needed to firmly verify and quantify the effect of BH spin-axis tossing, as well as for deriving constraints on the spin magnitudes of the first- and second-born BHs more robustly. 
    
    \item It remains a riddle to explain the physical mechanism of the tossing itself --- a process that may be generic to all core collapse SNe (NSs and BHs). 
    
    \item Producing BH+BH mergers with $\chi_{\rm eff}<0$ from isolated binaries {\em without} BH tossing requires a BH SN kick of order $w\sim 500\;{\rm km\,s}^{-1}$, in strong disagreement with observations of BHs in X-ray binaries.
    This finding is in agreement with previous work \citep[e.g.][]{wgo+18}.
    
    \item The crucial question of the efficiency of alignment of the spin axes prior to the second SN is still an unresolved issue.
    
\end{itemize}

\section*{Acknowledgements}
T.M.T. cordially thanks Thomas Janka for numerous discussions and explanations, Tom Callister for providing the PDFs of $\chi_{\rm eff}$ data from GWTC-3, Aleksandra Olejak, Simon Stevenson, Norbert Langer, Norbert Wex and Alejandro Vigna-Gómez for discussions, and Malte Klockmann Pein Malmberg for discussions during his BSc project on SN kinematics. Finally, an acknowledgement is given to the two anonymous reviewers and the scientific editor for ApJ, Ilya Mandel. 

\appendix
\section{Empirical data}\label{appendix:A}
The observational data used in this work for comparison with theory were taken from the recent GWTC-3 catalogue \citep{aaa+22a,aaa+22b} (\url{https://www.gw-openscience.org/eventapi/html/GWTC/}). 
The merger events which are double NSs (NS+NS: GW170817 and GW190425), putative mixed systems (BH+NS: GW190426\_152155, GW190917, GW191219 and GW200115) or BH+BH binaries with a secondary component mass that was most likely less than $3.0\;M_\odot$ (GW190814 and GW200210), were removed from the sample. This leaves a total of 85~BH+BH merger events. 
Since the uncertainty on the measurements of $\chi_{\rm eff}$ covers a range from $|\Delta \chi_{\rm eff}| < 0.1$ (GW191204) to $|\Delta \chi_{\rm eff}| > 0.4$ (GW200322), and typically $|\Delta \chi_{\rm eff}|=0.1-0.25$, the {\em individual} values of $\chi_{\rm eff}$ should be taken with a grain of salt, whereas the shape of the $\chi_{\rm eff}$ distribution for the {\em general} population is reliable.  
Selection effects \citep{gbo+18} from masses or spins, or internal relations \citep{chn+21}, are not considered here and are not expected to influence the main conclusions of this work, but see e.g. \citet{ms21} for discussions on priors used in the LIGO–Virgo–KAGRA analysis and the correlation between spin and mass ratio.

\bibliography{book_refs.bib}{}
\bibliographystyle{aasjournal}

\end{document}